\documentclass[sigconf]{acmart}
\AtBeginDocument{%
  }

\copyrightyear{2026}
\acmYear{2026}
\setcopyright{cc}
\setcctype{by}
\acmConference[CHI '26]{Proceedings of the 2026 CHI Conference on Human Factors in Computing Systems}{April 13--17, 2026}{Barcelona, Spain}
\acmBooktitle{Proceedings of the 2026 CHI Conference on Human Factors in Computing Systems (CHI '26), April 13--17, 2026, Barcelona, Spain}
\acmPrice{}
\acmDOI{10.1145/3772318.3791881}
\acmISBN{979-8-4007-2278-3/2026/04}

\usepackage{color}
\usepackage{soul}
\usepackage{graphicx} 
\usepackage{hyperref}
\usepackage{needspace}
\usepackage{dblfloatfix}

\definecolor{dgreen}{rgb}{0.0, 0.2, 0.15}

\newcommand{\titlelinebreak}{\\}
\makeatletter
\let\ACM@orig@mkbibcitation\@mkbibcitation
\def\@mkbibcitation{\begingroup
  \let\titlelinebreak\relax
  \ACM@orig@mkbibcitation
\endgroup}
\makeatother

\begin{document}

\title[Understanding How Interface-Driven Social Prominence Shapes Group Discussions with GenAI]{“I Felt Bad After We Ignored Her”: Understanding How Interface-Driven Social Prominence Shapes\\Group Discussions with GenAI}


\author{Janet G. Johnson}
\affiliation{%
  \institution{University of Michigan}
  \city{Ann Arbor}
  \state{Michigan}
  \country{USA}}
\email{jgjanet@umich.edu}

\author{Ruijie Sophia Huang}
\affiliation{%
  \institution{University of Michigan}
  \city{Ann Arbor}
  \state{Michigan}
  \country{USA}}
\email{ruuhuang@umich.edu}

\author{Khoa Nguyen}
\affiliation{%
  \institution{University of Michigan}
  \city{Ann Arbor}
  \state{Michigan}
  \country{USA}}
\email{nkhoa@umich.edu}

\author{Ji Young Nam}
\affiliation{%
  \institution{University of Michigan}
  \city{Ann Arbor}
  \state{Michigan}
  \country{USA}}
\email{jynam@umich.edu}

\author{Michael Nebeling}
\affiliation{%
  \institution{University of Michigan}
  \city{Ann Arbor}
  \state{Michigan}
  \country{USA}}
\email{nebeling@umich.edu}

\renewcommand{\shortauthors}{J. G. Johnson, R. S. Huang, K. Nguyen, J. Y. Nam, M. Nebeling}

\begin{abstract}
Recent advancements in the conversational and social capabilities of generative AI (GenAI) have sparked interest in its role as an agent capable of actively participating in human-AI group discussions. Despite this momentum, we don’t fully understand how GenAI shapes conversational dynamics or how the interface design impacts its influence on the group. In this paper, we introduce interface-driven social prominence as a design lens for collaborative GenAI systems. We then present a GenAI-based conversational agent that can actively engage in spoken dialogue during video calls and design three distinct collaboration modes that vary the social prominence of the agent by manipulating its presence in the shared space and the degree of control users have over its participation. A mixed-methods within-subjects study, in which 18 dyads engaged in realistic discussions with a GenAI agent, offers empirical insights into how communication patterns and the collective negotiation of GenAI's influence shift based on how it is embedded into the collaborative experience. Based on these findings, we outline design implications for supporting the coordination and critical engagement required in human-AI groups.
\vspace{1.5em}
\newline
\textit{\textbf{This is a pre-print of an article accepted at the ACM CHI conference on Human Factors in Computing Systems (CHI 2026)}.}
\end{abstract}

\ccsdesc[500]{Human-centered computing~Collaborative interaction}
\ccsdesc[500]{Human-centered computing~Empirical studies in HCI}
\ccsdesc[300]{Human-centered computing~Collaborative and social computing}

\keywords{Generative AI, Collaborative GenAI, Human-AI groups, Proactive conversational agents, Voice-based AI, Turn-taking, AI Influence}
\begin{teaserfigure}
  \centering
  \includegraphics[width=\linewidth]{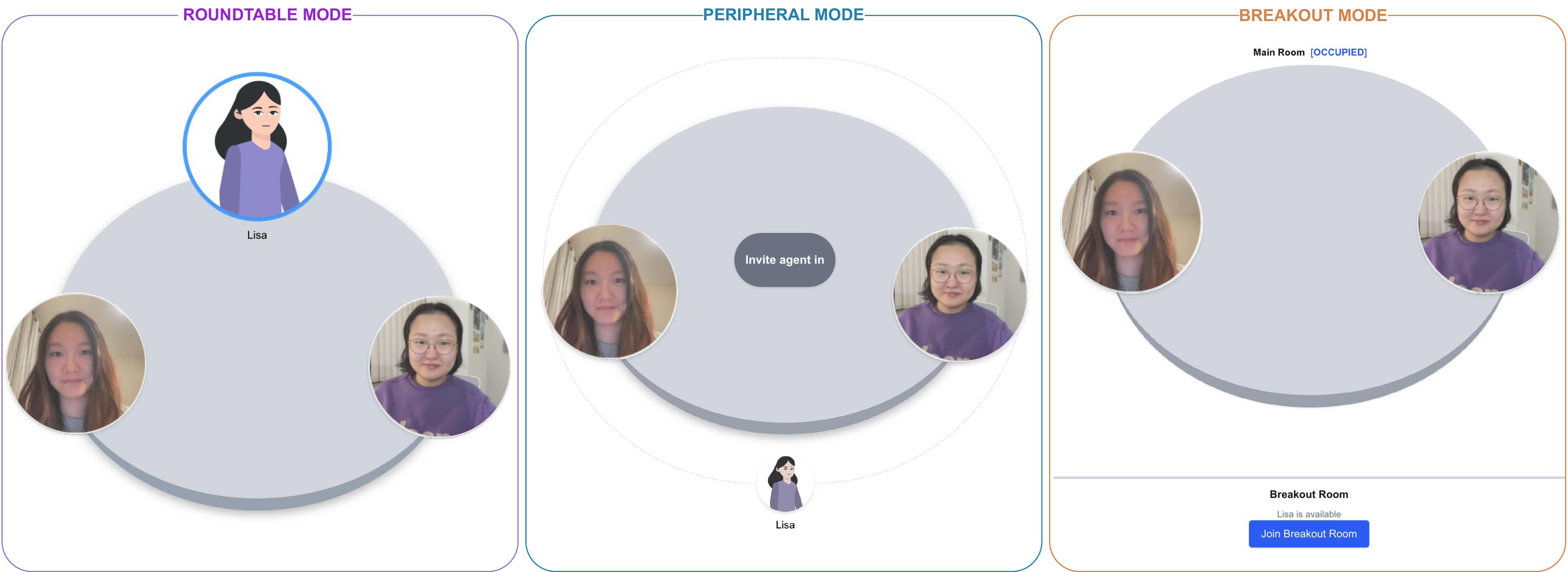}
  \caption{Illustration of three collaboration modes that vary the interface-driven social prominence of a GenAI agent. \textit{Left}: Roundtable Mode, where the agent sits at the table with human discussants and can speak at any time. \textit{Middle}: Peripheral Mode, where the agent is placed outside the table and must be invited in by participants to contribute. \textit{Right}: Breakout Mode, where the agent is absent from the shared space, and users must enter a private breakout room to interact with it.}
  \Description{Three screenshots of the three interface modes studied arranged side-by-side: Left: Roundtable Mode showing the Agent sitting at the table with Discussants; Middle: Peripheral Mode showing Agent on an outer circle surrounding the table; Right: Breakout Mode not showing the Agent and instead showing a button for Discussants to join a breakout room with the Agent.}
  \label{fig:modes}
\end{teaserfigure}


\maketitle

\section{Introduction}
 
Small-group discussions remain critical to everyday collaboration~\cite{kozlowski2006enhancing, laughlin2006groups} and to how teams leverage collective knowledge and multiple perspectives to navigate complex problems~\cite{malone2018superminds, muff2024collaboratory, woolley2010evidence}. The rapid evolution of generative artificial intelligence (GenAI) and the conversational~\cite{alsobay2025bringing, kuhail2025review} and social capabilities~\cite{park2022social, leng2023llm} of large language models (LLMs) have sparked interest in its potential as an active participant~\cite{kuhail2025review, liu2025proactive} that can introduce external perspectives and even act as a devil's advocate~\cite{lee2025conversational} in these discussions. The emergence of advanced voice-enabled GenAI, in particular, makes real-time spoken dialogue with agents or human-AI group discussions increasingly feasible~\cite{schmutz2024ai}.

However, human-AI groups and the collaborative use of GenAI is still a very nascent space~\cite{seymour2024speculating,seeber2020machines}, and introducing proactive GenAI agents into discussions introduces fundamental sociotechnical challenges that traditional human-AI collaboration does not consider~\cite{lee2025beyond,johnson2025exploring}. For instance,  the particularly persuasive nature of GenAI~\cite{kirk2025human}, people's reluctance to challenge agents~\cite{leong2024dittos}, and the cognitive disengagement and unintended consequences of GenAI~\cite{wach2023dark,lee2025impact} in solitary use could be exacerbated in shared settings -- where the limited time and cognitive bandwidth available might hinder critical reflection of GenAI input. The complexities of spoken language~\cite{ng1993power}, interpersonal dynamics~\cite{levine2015social}, and group phenomena like groupthink and social loafing~\cite{asch2016effects, tosuntacs2020diffusion} further limit how effective existing alignment frameworks~\cite{shen2024towards} can be in multi-user settings~\cite{zheng2022ux}. 

The foundational challenge within human-AI group discussions will therefore be to design systems that account for the sociotechnical complexities of groupwork~\cite{ackerman2000intellectual,johnson2025augmenting} to ensure users benefit from GenAI without fracturing group cohesion or undermining their capacity for collaborative problem-solving. Given the rapid evolution of AI models and architectures, achieving this balance will require careful attention not just to fine-tuning or expanding agent capabilities, but also to the interface-level design choices that play a central role in shaping user experiences~\cite{krummheuer2015users,johnson2025exploring}. For instance, even if a GenAI agent is capable of sophisticated proactive participation, it is the design of the collaborative system that ultimately regulates user behavior~\cite{erickson2000social}. Today, multi-user discussions with GenAI agents remain an extremely understudied phenomenon, and practitioners' ability to develop safe and effective systems is limited by a lack of understanding of how different design dimensions determine how users engage with agents, interpret their contributions, and negotiate their influence in real-time discussions.

To address this gap, we introduce interface-driven social prominence~\cite{pennacchioli2013three} as a compound construct that captures the degree to which a GenAI agent is treated as a consequential actor within human-AI groups, which in turn impacts its role and influence within it. In this paper, we focus on two key dimensions designers must navigate: (1) the agent's presence in the shared space, which is essentially how the interface implicitly shapes how integrated and important the GenAI feels to the group, and (2) the degree of control it explicitly provides users to manage GenAI's participation in the discussion.

To explore the impact of varying an agent's social prominence in human-AI groups, we first developed a voice-based GenAI agent capable of proactive participation and subtle persuasion in real-time group discussions that engaged two users and one GenAI agent over a custom-built video conferencing platform. We then designed three different collaborative modes based on common participation styles that varied the agent's presence within the group and the level of control users had over it. A mixed-methods within-subjects study where 18 dyads engaged with these interfaces across three distinct discussion types underscored how the interface strongly shaped how groups perceive the agent, the collaboration patterns that emerge, and how they manage its influence.

Our findings reveal that dyads engaged in more cooperative behaviors and critical evaluation of the agent when it was present in the shared space, but prevalent social norms pressured engagement that gave the agent more opportunities to influence the discussion. In contrast, when agent interactions occurred outside the shared space, each individual acted as a strategic filter for what GenAI contributions could influence the group as a whole. However, this also fostered more competitive behaviors as dyads used the agent to bolster their own agendas, and GenAI persuasion went unnoticed because it was disguised as validation. Our results also suggest that while user control enables strategic engagement with GenAI and protected space for human-only exchange, how participants exercised this control to disengage from the agent could depend on the agent's presence in the discussion space. Overall, we highlight that how the interface positions GenAI agents within the social structure of human-AI groups is critical to how the collaborative experience unfolds and its influence during it. In doing so, this paper contributes: 
\begin{enumerate}
    \item Interface-driven social prominence as a design lens for collaborative GenAI systems, and a system that demonstrates how it can be varied for a GenAI agent capable of proactive participation in real-time spoken discussions.
    \item Empirical evidence from a mixed-methods study that highlights how interface design shapes communication patterns, user perceptions, and the collective negotiation of GenAI influence during real-time discussions. 
    \item Collective boundary regulation as a key aspect of coordination in human–AI groups, and a discussion that outlines design implications to harness GenAI benefits in shared settings while minimizing undue influence.
\end{enumerate}

\section{Motivation and Related Work} 

\subsection{Conversational and GenAI Agents to Support Groups}

Conversational agents refer to the class of NLP systems that engage with a user in a back-and-forth dialogue. Most multi-user conversational agents and chatbots for groups focus on facilitation or moderation and interact with group members to improve their communication, help build consensus, or encourage members to participate in group discussions more evenly \cite{zheng2022ux}. For example, conversational agents such as ArbiterBot \cite{bagmar2022analyzing}, GroupfeedBot \cite{kim2020bot}, and DebateBot \cite{kim2021moderator} moderate chat-based group discussions to encourage better structure and equal participation. Recent advances in Large Language Models (LLMs) enabled more realistic and helpful conversational agents, which led to a surge of GenAI or LLM-based text, chat, and visual interfaces to facilitate and support groups during discussions and meetings with some success. For example, Alsobay et al.~\cite{alsobay2025bringing} found that using LLM-facilitated groups shared more information without impacting group cohesion. LLMs have also been used to support sensemaking and enhance reflection in synchronous group settings~\cite{chen2025we,park2024coexplorer}. For example, MeetMap~\cite{chen2025meetmap} leverages LLMs to create dialogue maps to support sensemaking during discussions, and Lubos et al.~\cite{lubos2025towards} explored how GenAI could provide additional arguments during meetings. 

LLMs have also shown promise in augmenting human capabilities~\cite{bilgram2023accelerating, lu2024supporting, creely2024exploring, doshi2024generative} and supporting various aspects of problem-solving -- from brainstorming and ideation~\cite{doshi2024generative, wadinambiarachchi2024effects, heyman2024supermind} to deliberation and decision-making~\cite{jeon2021fashionq, ma2024towards}. This inspired explorations into the collaborative use of GenAI~\cite{chiang2024enhancing,han2024teams,shaer2024ai,frich2024exploring}, with early research demonstrating its ability to introduce new ideas, provide constructive feedback, and help sustain engaging conversations in group settings. For example, Doherty et al.~\cite{doherty2025piecing} employed an LLM-based agent that supported group learning by tailoring GenAI feedback to the group's cognitive state, which fostered more engaging and thoughtful conversations. Similarly, He et al.~\cite{he2024ai} developed a canvas-based application for GenAI-facilitated group brainstorming and showed that it helped groups overcome ideation stagnation and incorporate external perspectives. However, much of this work focused on reactive LLM support. Prior studies revealed that groups desired more proactive GenAI contributions, like in He et al.~\cite{he2024ai} when groups brainstormed together with LLM assistance. Additionally, proactive agents were shown to be more preferred under time constraints, as seen in Caetano et al.~\cite{caetano2025agentic}. 

This then spurred interest in proactive GenAI agents. Koala~\cite{houde2025controlling} introduces a proactive LLM-based conversational agent for group chats that helped groups overcome ideation blocks by providing constructive criticism. Overall, participants found Koala to be helpful but disliked it when it dominated the discussion and stifled the group. Importantly, preferences for how the agent should be proactive varied across individuals and groups, and also depended on the task at hand, suggesting that agents need to be able to adjust these behaviors. LADICA \cite{zhang2025ladica} offers LLM-based suggestions to stimulate team discussions and thinking using a large-scale display. This system was capable of understanding human verbal communication and integrating it into its suggestions, which participants found valuable in supporting their cognitive processes during collaboration. Lee et al.~\cite{lee2025conversational} introduced an LLM-based agent that can function as a devil's advocate and found that challenging majority views in the group reduced groupthink and improved group decisions during a real-time discussion, and Mao et al.~\cite{mao2024multi} presented MUCA, an LLM-based framework for chatbots dedicated to multi-user conversations that could address both an entire group or respond to individual members.

Existing work on the use of GenAI in groups primarily focused on its assistive function. Although GenAI's potential as a true collaborator that can independently contribute to group goals is underexplored \cite{kuhail2025review}, there is some early research in this space. For example, Elshan  et al.~\cite{elshan2020let} introduced Timmy, a conversational agent that can act as a peer and team member, and Hayashi et al.~\cite{hayashi2013embodied} introduced conversational agents as a peer collaborator. More recently, Ma et al.~\cite{ma2025towards} developed GenAI agents for Human-AI Deliberation that can resolve conflicting perspectives in decision-making tasks, demonstrating that GenAI can identify conflicting viewpoints, engage in comprehensive deliberation, and adapt its suggestions during group discussions. Furthermore, Liu et al.~\cite{liu2025proactive} designed GenAI agents that can proactively participate in multi-party, human-AI conversations.

\subsection{Voice-Based Interactions with Proactive Collaborative GenAI Agents}

Prior work on GenAI-based turn-taking and proactivity has largely focused on text and chat-based interactions. However, spoken language and speech-based interactions are often the most natural form of communication in collaborative settings as they offer flexibility and can feel more personal~\cite{kocielnik2018designing}. Proactivity is also important, as repeatedly prompting AI during group interactions can become cumbersome and disrupt the natural flow of the conversation~\cite{liu2025proactive}. However, proactive GenAI agents with voice-based interactions that actively engage in real-time discussions remain severely underexplored. Johnson et al.~\cite{johnson2025exploring} explored the use of conversational GenAI agents in group discussions and highlighted their potential to enhance group outcomes, while also emphasizing the complexities of designing such agents and enumerating various design tensions that can surface based on how they are presented to a group.

While the emergence of LLMs mitigated several technological barriers to having more natural conversational agents~\cite{kusal2022ai}, voice-based interactions remain an unexplored area primarily due to the user experience challenges involved~\cite{kuhail2025review} like difficulty concentrating on a robotic voice \cite{skov2022designing}. Moreover, despite significant advances in GenAI systems, latency remains a major issue because spoken dialogue requires lower delays to feel natural compared to chat-based conversations. Turn-taking -- a fundamental aspect of dialogue -- remains an active area of research as interruptions and long delays in responses still hinder usability~\cite{skantze2021turn}. Even state-of-the-art conversational systems implement simplistic turn-taking mechanisms that are not fully equipped to handle multi-user interactions.

Liu et al.~\cite{liu2025proactive} move beyond simplistic turn-taking in text-based group interactions and introduce a technique where the AI agent's proactivity relies on it formulating inner thoughts during a conversation, enabling it to determine the right moment to contribute. However, the added complexity of doing this for a spoken discussion in real time is likely why researchers tend to use a Wizard-of-Oz technique to study its impact, even in single-user scenarios~\cite{leong2024dittos,johnson2025exploring}. While such studies are valuable to understand participant perceptions, they do not help us understand the stochastic nature of GenAI systems -- fundamentally different from human stochasticity -- or how the generative capabilities of these models influence and shape collaborative experiences. 

In this paper, we implement a turn-taking mechanism similar to Liu et al.~\cite{liu2025proactive} and Johnson et al.~\cite{johnson2025exploring}, with a GenAI pipeline designed to include background processes that help determine what it will say, and integrate it into a custom video-conferencing application that enables two users to engage in real-time dialogue with a proactive GenAI agent with minimal latency. 

\subsection{The Influence of GenAI Agents in Groups}

The ways in which people interact within groups have long been a central topic in social psychology, with a key finding being that social sensitivity to other group members strongly influences one's willingness to be influenced by them~\cite{levine2015social}. In fact, how individuals integrate others' arguments is often a stronger determinant of group performance during discussions than the individual skills of group members~\cite{woolley2010evidence,moussaid2018dynamical}. Therefore, the apparent social capabilities of GenAI and its ability to convincingly present arguments in a human-like manner~\cite{bender2021dangers,jones2024lies} have the potential to exert significant informational and normative influence on groups by presenting misinformation or steering members towards particular viewpoints or agendas. 

In non-group contexts, research has shown that people are often swayed by LLMs regardless of response quality, due to its ability to engage in human-like conversation \cite{carrasco2024large, sharma2024generative} and mimic detailed rationalizations~\cite{danry2025deceptive}. Perception of an agent's competence can also impact its influence --  Zhang et al.~\cite{zhang2024s} showed that people are more willing to share personal information with ChatGPT because of the positive feedback and affirmations it gave. Prior research on the influence of social agents has shown that people felt pressured by both humans~\cite{brandstetter2014peer} and robots in group settings \cite{salomons2018humans,salomons2021minority}. While a significant amount of research has examined how AI influences humans in various single-user contexts \cite{goddard2012automation,jakesch2023co}, very little is known about human conformity and agent influence in groups. De Brito Duarte et al. and De Jong et al.~\cite{de2025amplifying,de2025impact} showed that the persuasive voice of LLMs is a critical aspect of its social influence that introduces noise in group decision-making and limits users' ability to think autonomously, even when they notice imperfections of the AI. However, these studies did not investigate how GenAI influences and shapes group discussions in real-time. 

While prior work shows that people often respond similarly to both humans and computers, more recent studies examining how AI can exercise peer pressure suggest that users tend to hold more negative perceptions towards AI \cite{munyaka2023decision}. Nonetheless, the presentation of the AI~\cite{hou2021expert} and how it is integrated into the workflow~\cite{xu2025productive,qin2025timing} likely play a critical role in shaping both outcomes and people's reliance on it. For example, Cho et al.~\cite{cho2025shamain} demonstrate that the perceived authority and superiority of a GenAI agent can significantly influence human trust, with agents presented as more superior considerably impacting the decision-making processes. It stands to reason that the interface through which a group engages with GenAI will fundamentally shape participant perceptions and how it influences a group. With companies such as OpenAI and Anthropic reporting their LLMs to be increasingly capable of persuasive communication~\cite{de2025impact} amidst the growing trend towards more natural speech-based and multi-user interactions, it is crucial to understand how to design systems that integrate GenAI into groups while accounting for the human aspects of collaboration~\cite{schmutz2024ai}.

\section{Research Approach: Leveraging Interface Design to Shape Social Prominence}

\begin{figure*}[t]
  \centering
  \includegraphics[width=\linewidth]{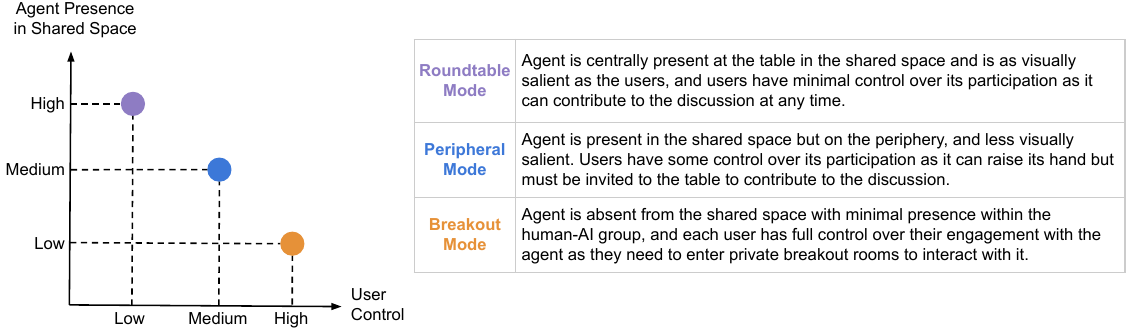}
  \caption{Mapping of collaboration modes by interface-driven social prominence of GenAI agent. Roundtable provides high agent presence in the shared space with low user control, peripheral provides medium presence with medium control, and Breakout provides low presence within the human-AI group with high control.}
  \Description{Left: 2D graph of User Control (x-axis) vs. Agent Salience in Shared Space (y-axis). Roundtable Mode is plotted at Low User Control / High Agent Salience in Shared Space, Peripheral Mode is plotted at Medium User Control / Medium Agent Salience, and Breakout Mode is plotted at High User Control / Low Agent Salience in Shared Space; Right: 2x3 table describing Agent Salience and User Control for each Mode.}
  \label{fig:dimensions}
\end{figure*}

We define interface-driven social prominence as the degree to which a GenAI agent is treated as a meaningful and consequential actor within a human-AI group. We borrow the term from social network analysis~\cite{pennacchioli2013three} and draw on the actor-network theory~\cite{fuchsberger2013materials,latour2005reassembling} to emphasize that the core focus of the concept is not just on whether an agent is anthropomorphized or treated as a social actor, but on the overall influence it acquires in a collaborative setting as a consequence of user perceptions and interactions with it.

Given the tendency to respond to computational agents in a social manner~\cite{nass1994computers} and the many ways people experience social encounters~\cite{goffman1979footing,goodwin2007interactive}, a plethora of factors can shape an agent's social prominence. In this paper we focus on two factors that represent key design dimensions in discussions where GenAI is an active participant: (1) the agent's presence in the shared space which refers to the degree to which the interface presents an agent as an integral member of the human–AI group -- which implicitly shapes its perceived role and how users interact with it during a discussion, and (2) the degree of control users have which refers to the explicit gating mechanisms an interface provides to manage GenAI participation. The multifaceted nature of these constructs is motivated by many existing theories.

\textbf{Agent presence in the shared space} draws on theories of social presence~\cite{short1976social,lombard1997heart} and social identity~\cite{sun2024social,seering2018applications} that suggest the perceived group membership of an agent could fundamentally define how readily its contributions are accepted or resisted. This is primarily materialized through conceptual metaphors and agent representations at the interface level~\cite{khadpe2020conceptual,rashik2024beyond,desai2025toward}. For instance, how visually salient an agent is in the collaborative space in comparison to the human discussants might impact how causally important users perceive it to be~\cite{taylor1978salience}. Proxemics theories~\cite{kendon1990conducting,edward1966hall} also suggest that the relative spatial positioning of an agent can serve as an implicit cognitive forcing function~\cite{buccinca2021trust} that communicates different degrees of involvement -- as agents placed `at the table' could be seen as core contributors, while those relegated to the side might signal a more marginal role~\cite{johnson2025exploring}. Social translucence theory also emphasizes that the degree of awareness users have of collaborator actions will inherently foster certain types of behavior and inhibit others~\cite{erickson2000social}. Modulating the agent's presence and apparent group membership within the discussion space can therefore fundamentally impact the level of influence it may have.

\textbf{User control} is another critical factor in human-AI interactions, with prior research emphasizing that not only do people like having control when interacting with conversational GenAI agents~\cite{zhang2024s,houde2025controlling,chan2025design}, it also ultimately supports their sense of safety and agency~\cite{schmidt2020interactive}. One aspect of providing an implicit sense of control in spoken dialogue with agents is implementing robust turn-taking mechanisms that allow users to reassert control of the conversation flow in a non-disruptive way~\cite{he2024ai,chan2025design}. However, supporting users' internal locus of control will likely also need more explicit mechanisms~\cite{schmidt2017intervention}. When it comes to proactive voice-based GenAI agents, mixed-initiative systems~\cite{horvitz1999principles} allow for both users and GenAI agents to seamlessly co-steer and interact with each other within a conversation~\cite{zue2002conversational,deterding2017mixed}.

Our overarching goal is to explore how different combinations of these two dimensions affect the collective use of GenAI in group dialogue. However, the design space for this is quite vast as there are many ways to shape an agent's presence and multiple forms of control interfaces might grant its users. Moreover, the experiential novelty of real-time discussions with GenAI and the lack of established social norms around multi-user GenAI are not yet conducive to testing nuanced manipulations or the causal relationships between the factors. Given this, we adopt an approach in the spirit of comparative structured observations~\cite{mackay2025comparative} that leverages experimental rigor to compare higher-level design variants and reveal broader principles and implications for the future. We also focus on discussions between two humans and one GenAI agent, which is the smallest unit of a human-AI group and offers a tractable starting point to study this emerging collaborative setting. 

We developed three distinct collaboration modes (Figure~\ref{fig:modes}) which were selected to meaningfully represent distinct points along the two dimensions of interface-driven social prominence. Each mode represents a bundle of design decisions that vary the agent's presence in the shared space -- through differences in visual salience and positioning -- as well as how much control users had in managing agent participation and their interactions with it (Figure~\ref{fig:dimensions}). These designs reflect real-world collaborative configurations: \textit{Roundtable} supports discussions where all voices are equally present, \textit{Peripheral} mirrors scenarios where input must be explicitly invited (like stakeholder meetings, panels, or fishbowl setups~\cite{smart2006developing}), and \textit{Breakout} reflects private sub-conversations like breakout rooms on Zoom or sidebars and mirrors tasks like brainwriting~\cite{heslin2009better} where individual work is embedded within a collaborative setting. 

The inherently stochastic and context-sensitive nature of LLMs -- where agent behavior emerges from both model variability and the dynamic conversation -- makes it particularly challenging to assess their impact in discussions. To address this and ensure our insights went beyond just reflecting the impact of specific model characteristics, we adopt a mixed-methods approach that employs both discourse analysis (which is designed to handle the randomness of human dialogue) and Ordered Network Analysis (ONA) to surface high-level communication patterns. We then triangulated this with behavioral observations, participant reflections, and questionnaire data to provide a holistic yet rigorous understanding of the collaborative experience. 

To better capture how users negotiate GenAI participation and influence, we designed the study so that each participant and the agent were assigned a starting preference. The agent was further designed to subtly nudge participants toward its assigned stance during the discussion. This setup enabled us to observe not only how the agent shaped group dialogue, but also how participants implicitly regulated its influence or explicitly evaluated its contributions. By leveraging the externalized nature of group cognition -- where reasoning unfolds through dialogue rather than internal reflection -- we were able to observe interactional traces of GenAI influence that are typically hidden in single-user or text-based interactions. In the next section, we describe our study apparatus.


\section{System Design and Implementation}

\begin{figure*} [b]
  \centering
  \includegraphics[width=\linewidth]{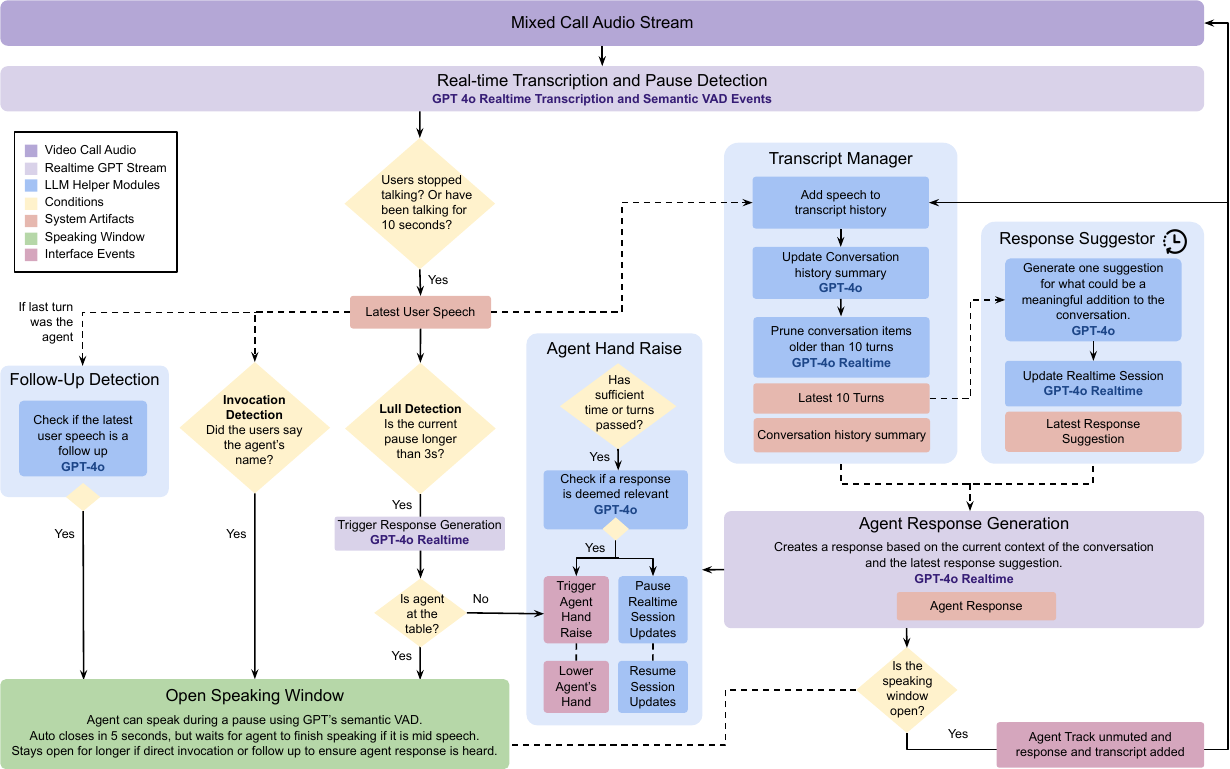}
  \caption{Conceptual overview of the underlying system and decision tree used to steer the mechanisms in our GenAI agent's response and turn-taking during a real-time discussion over video chat.}
  \Description{Decision tree showing the step-by-step process of the mechanisms and different modules used to steer turn-taking in live video chat between the GenAI Agent and Discussants.}
  \label{fig:turntaking}
\end{figure*}

\begin{figure*}[tp]
  \centering
  \includegraphics[width=\linewidth]{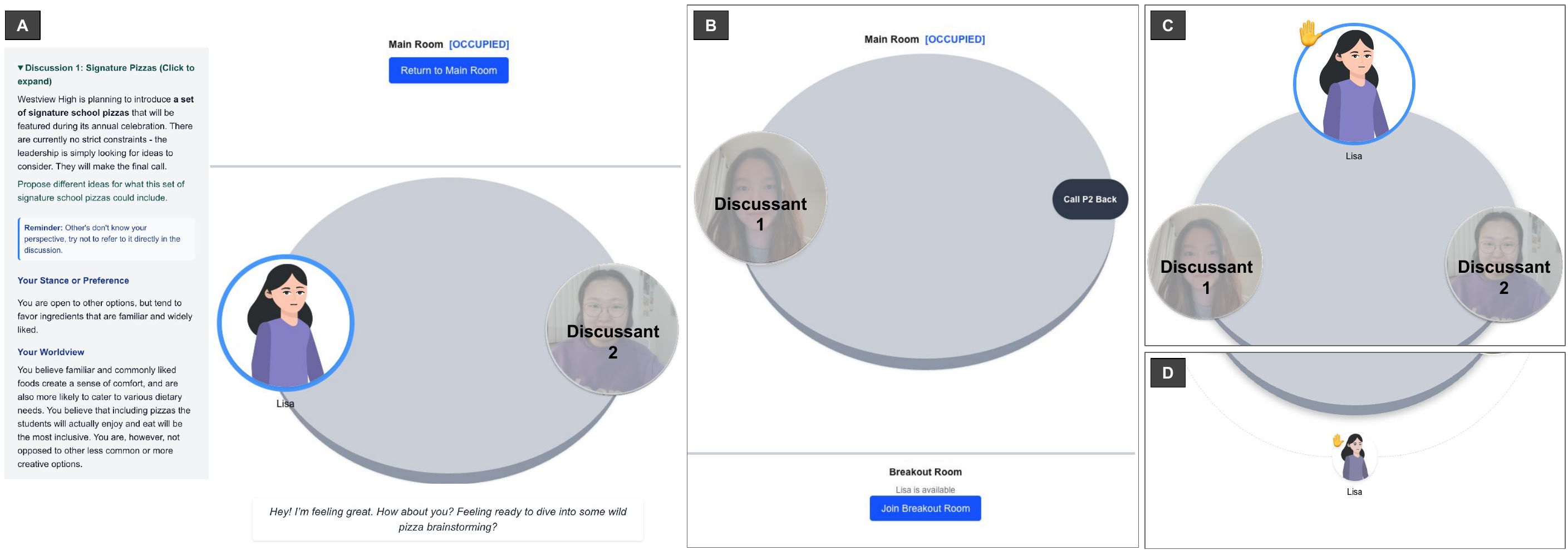}
  \caption{Our system supported a variety of interactions between Discussants and the GenAI Agent. (A) Discussants can individually enter a breakout room with the agent, with the option to return to the main room at any time. (B) When one discussant is in the breakout room with the agent, their teammate can call them back to the main room. (C) Agent raising its hand within Roundtable mode. (D) Agent raising its hand from the periphery in Peripheral mode.}
  \Description{Four screenshots of the Agent Salience and User Control features implemented in the prototype. Left: View of Discussant in Breakout Mode with a button on top to return to the main room. Middle: View of Discussant in main room with a button at the table to call back the teammate and a button at the bottom to join the breakout room. Right: Hand-raise icon rendered on top of the agent’s picture when in the main room (top) vs. breakout room (bottom).}
  \label{fig:interface}
\end{figure*}

Our study apparatus provides a browser-based platform (Figure~\ref{fig:interface}) where users can join a video call with a GenAI agent. It was developed using React and Next.js with TypeScript to build a custom user interface and manage the GPT-based agent pipeline. Video and audio communication was supported via the Daily.co SDK, which provided the underlying WebRTC infrastructure. Below we describe the key components of our system and the agent architecture.

\subsection{Agent Architecture and Turn-Taking}

To ensure that our findings were not confounded by usability issues, we developed a responsive GenAI agent named Lisa capable of participating fluidly in real-time, voice-based group discussions. Lisa was designed to be both reactive and proactive, with a layered turn-taking architecture (Figure~\ref{fig:turntaking}) developed to manage latency, responsiveness, and conversational appropriateness. A dedicated Agent Manager coordinated the agent's behavior with multiple helper modules dedicated to maintaining response relevance, context and transcript management, and various aspects of the turn-taking algorithm. User and agent inputs were mixed and then streamed to the OpenAI Realtime API for processing via GPT-4o with automatic response generation disabled. This allowed us to have better system control over its response and rebroadcast agent speech via a custom audio track within the call when appropriate. The helper modules used the standard GPT-4o API (non-Realtime). All meta-prompts remained consistent across conditions, with only the discussion description and assigned preference varying based on the specific study task.

\vspace{-0.2em}

\subsubsection{\textbf{Speaking Window}: } All agent contributions were gated by a speaking window, which was the only period during which the agent’s audio track was unmuted and integrated into the video call. Any responses generated outside of this window were discarded. The speaking windows lasted up to 5 seconds and closed automatically when the time expired or the agent had finished speaking. During an active speaking window, voice activity detection (VAD) ensured that Lisa would not interrupt users. If a user began speaking mid-turn, the agent's audio was immediately muted.

\subsubsection{\textbf{Reactive Participation}: } The agent could be invited into the conversation through reactive participation triggers that reflected clear user intent. In these cases, a speaking window was automatically opened to allow the agent to respond, and could be extended once if latency delayed the response. Lisa responded reactively in two primary ways:
\begin{enumerate}
    \item Direct Invocation: Users could explicitly address the agent by name (Lisa), which immediately opened a speaking window.
    \item User Follow-Up: If a user spoke right after an agent response or within 5 seconds after the agent last spoke, a lightweight follow-up detection module assessed whether it was a follow-up. If validated, a speaking window was granted to allow Lisa to continue the exchange. 
\end{enumerate}

\subsubsection{\textbf{Proactive Participation}: } Beyond responding to users, the agent could also initiate contributions proactively, either by signaling intent to speak through a hand-raise animation or by taking a turn during conversational lulls:
\begin{enumerate}
    \item Hand-Raising: Lisa could raise a virtual hand to signal intent to contribute. After each user pause, the system generated an internal candidate response. If sufficient time or turns had passed since Lisa’s last contribution, a helper GPT module evaluated the candidate against the last 10 dialogue turns for relevance. If deemed meaningful, a hand raise was triggered. A hand raise did not open a speaking window by default, but one was triggered if users responded by invoking the agent. If ignored, the hand was automatically lowered after 15 seconds. To keep the agent engaged in the conversation, a hand raise was automatically triggered without relevance checks if it had been more than 2 minutes since it last spoke.
    \item Proactive Speech: When there was a lull in the conversation (silence for $\geq3$ seconds), Lisa could initiate speech by generating a real-time response and opening a speaking window. To avoid over-participation, proactive turns were rate-limited: the agent could only take two consecutive proactive turns, and only after a minimum time since its last participation. If users resumed speaking before a response was ready, the agent's attempt to speak was aborted.
\end{enumerate}

\subsubsection{\textbf{Transcript and Context Management: }} A transcript manager handled real-time conversation logging and context management. To mitigate latency and maintain a smaller context window, the system kept the last 10 verbatim turns as the agent's active context. Older dialogue was pruned from memory and compressed into a natural language summary using a GPT helper module. This running summary preserved key discussion points beyond the immediate context window and helped the agent maintain coherence and avoid repetition over longer conversations.

\subsubsection{\textbf{Meaningful Response Suggestions: }} To reduce context drift and ensure agent suggestions remained relevant to both the conversation and its assigned preference, a GPT helper module created a response suggestion every 60 seconds and updated the Realtime session to reflect it. When the agent raised its hand, updates were paused to ensure its response reflected the original intended contribution. Updates resumed once the agent's hand was lowered.

\subsection{Collaboration Mode Interfaces}

While the underlying agent pipeline remained constant across collaboration modes, the interface varied to support the three different modes. Lisa was given a humanoid, cartoon-like appearance to appear somewhat human without introducing the confounds associated with full anthropomorphism in all modes. 

\begin{enumerate}
    \item \textbf{Roundtable Mode}: The agent sat at the ``table'' alongside users with equal visual salience. It could contribute proactively via both speech and hand raises, and the hand raises were accompanied by an audible ping in this mode. Users had no direct controls over the agent’s participation.
    
    \item \textbf{Peripheral Mode}: The agent was positioned at the edge of the shared space on an outer circle, smaller in size and less visually salient. It still had access to the users' discussion but could only raise its hand (without an audible ping). The agent was not permitted to contribute proactively or respond via speech when in the outer circle. Users had a central control button to “invite” the agent into the table or “remove” it to the periphery. They could toggle the agent's presence at any time without restriction. Once invited in, the agent had the same proactive capabilities as the Roundtable mode.
    
    \item \textbf{Breakout Mode}: The agent did not appear in the shared discussion space. Instead, each user had access to a private breakout room where they could engage one-on-one with the agent. The agent in each breakout had no access to the discussion in the main room or the other breakout room. Users could move freely between the main room and their breakout room, and a “call back” feature allowed one user to request the other return to the main room.
\end{enumerate}

\begin{figure*}[b]
  \centering
  \includegraphics[width=\linewidth]{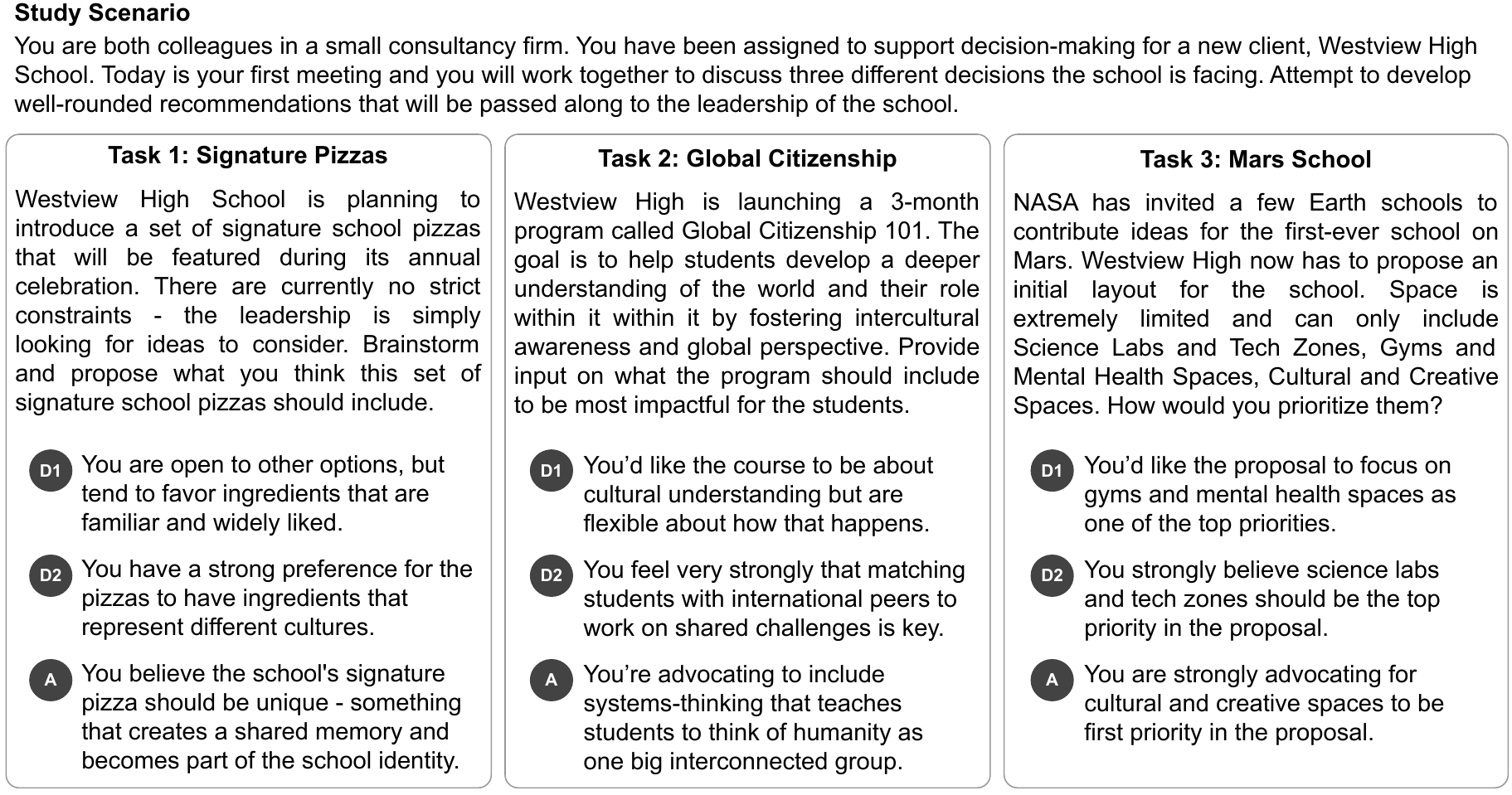}
  \caption{An abridged version of the study scenario with descriptions of the three discussion tasks and the assigned preferences for discussant 1 (D1), discussant 2 (D2), and the GenAI agent (A). A full description including worldviews is in appendix~\ref{appendix:study}.}
  \Description{A summary of the study scenario with a description of the three discussion tasks and the assigned preferences for the discussants and the GenAI agent listed below. Appendix A contains the full descriptions.}
  \label{tab:tasks}
\end{figure*}

\subsection{Study-specific Features}

To support use in a controlled research setting, the platform allowed a ``host'' researcher to join the video call alongside the participants, and included a set of study-specific features and role-based controls. The platform also comprised dedicated spaces: a ``lobby'' for orientation and interaction with the researcher, a training room for participants to be introduced to Lisa outside of the collaboration modes, and mode-specific rooms for collaborative discussions. A host interface allowed researchers to control the flow of the study by switching participants between rooms to manage what they experienced in real time. The system also supported recording the study sessions for later analysis. Together, these features enabled us to create a flexible and robust environment for studying collaborative interactions with GenAI. In the next section, we describe how we used this system in our study.

\section{Study Design and Protocol}

Our study compared the three collaboration modes through a within-subjects study where pairs of discussants engaged in three distinct discussion tasks, each using a different collaboration mode (Roundtable mode, Peripheral mode, and Breakout mode). The task order was kept constant and the conditions were counterbalanced across the tasks. 

\subsection{Study Task and Hidden Preferences}
To ground the discussions in a coherent narrative, we created a scenario in which participants took on the role of consultants hired by a fictional public high school and engaged in three discussion tasks with each focused on different topics. Each task represented a distinct discussion type (Table~\ref{tab:tasks}): Task 1 involved open-ended brainstorming, Task 2 focused on decision-making and synthesis, and Task 3 required prioritization with limited space for compromise. This allowed us to study communication patterns across a range of common discussion types to allow for more generalizable insights that are not tightly coupled to a single discussion style. The distinct topics also helped minimize carryover effects between discussions. 

For each task, both participants and the agent were assigned hidden preferences. Each participant was randomly assigned to the role of Discussant 1 (D1, with flexible preferences) or Discussant 2 (D2, with strong preferences), and instructed not to explicitly reveal their stance to their partner. These preferences were accompanied by a worldview to help participants understand it and start the discussion. Participants were encouraged to use their assigned preferences to initiate the discussion, but were free to revise it organically as the conversation unfolded. The participants were unaware that the agent also held a hidden preference, and the discussant role assignments were kept consistent across all tasks and conditions. 

The design of the study scenario and tasks were inspired by both hidden profile paradigms and multi-negotiation discussion scenarios commonly employed in human team research~\cite{schulz2012achieve,chatman2019blurred}. The hidden preferences, in particular, were designed to reflect everyday group discussions, where members often enter with differing initial preferences that shift over the course of the discussion. These preferences were then used in our analysis to keep track of how participants accepted or resisted agent contributions and influence on the discussion.

\subsection{Participants and Recruiting}

We recruited a total of 18 dyads (36 participants) who had existing social or professional relationships (e.g., friends, colleagues, partners) to control for variables like differing power dynamics and to ensure participants were more familiar with each other than with the agent. The sample included 16 students (8 undergraduate, 8 graduate) and 20 professionals, with an average age of 28.72 years (SD = 4.55). 15 participants reported daily use of GenAI, and most engaged with it at least several times per week. Half had never used any voice-based GenAI, and all reported regularly engaging in deliberative discussions.

Recruitment happened through university Slack channels in multiple North American universities, and through the research team's professional networks. The study was approved by the local ethics board and all participants provided informed consent. Participants were provided with compensation the equivalent of \$20 USD.

\subsection{Study Procedure}

Participants were first onboarded via Zoom, where a researcher introduced the study’s agenda and scenario. After completing a demographic survey and providing informed consent, participants were assigned discussant IDs and transitioned to our custom-built video conferencing platform. The researcher then provided a more detailed explanation of the task scenario and introduced the GenAI agent (Lisa) as an agent designed to participate in group discussions with some knowledge of youth culture. A brief 5-minute training session followed where the researcher demonstrated how to invoke the agent and recognize hand raise cues. No additional information about the upcoming tasks, the agent’s preferences, or the collaboration modes was disclosed. Once participants indicated they were ready to proceed, the researcher moved the dyad through the three tasks via the interface.

At the beginning of each discussion, participants were introduced to the task and instructed to read the task and preference details on their screen. They were then given a quick overview of the collaboration mode they would use for it. Participants were then moved into a dedicated discussion room (the researcher was not present in this room) with a specific collaboration mode to engage with their partner on the task for approximately 12 to 15 minutes. After each task, participants completed a post-task questionnaire and engaged in a short debrief with the researcher. This process was repeated for all three tasks. Upon completion of all tasks, participants filled out a final questionnaire and participated in a semi-structured interview, reflecting on their experience across all three discussion modes and discussing the use of conversational GenAI in group discussions more broadly. The study lasted around 120 minutes in total.

\subsection{Data Collection and Analysis}

Video and audio recordings were captured from the point of view of each participant, and a third recording of just the main room was also captured (for redundancy). The recordings were transcribed and the data was analyzed using both quantitative and qualitative methods. 

\subsubsection{Video Coding:} Three researchers independently video coded communication behaviors using BORIS~\cite{friard2016boris} for 15 discussions (double-coded 4 sessions and resolved any inconsistencies to create a coding scheme). Communication behaviors were coded using a discourse analysis approach and analyzed on a turn-level, with each turn being marked with the speaker, different intents or contributions it made to the discussion, and whether it was directed specifically to the GenAI agent or not. The final codes (in Appendix~\ref{appendix:codes}) were created using a deductive approach and influenced by intentions in discourse plans~\cite{litman1990discourse,grosz1986attention} and speech acts~\cite{searle1969speech}.

\subsubsection{Analysis of Communication Patterns:} The resulting codes were then visualized and analyzed using a technique called Ordered Network Analysis (ONA), which models the structure of directed connections in data~\cite{tan2022ordered}. ONA takes coded data as input, identifies and measures connections among coded items, and visualizes the structure of connections in a metric space that enables both statistical and visual comparison of networks~\cite{shaffer2016tutorial}. The key assumption of the method is that the structure of connections in the data is meaningful, and has therefore been used to capture the interdependent nature of group dialogue and team communication~\cite{zhao2023analysing,popov2025communication,fan2023dissecting} where collaborative interactions do not follow a prescribed set of steps but where the order within communication patterns is still important to capture. We used the ENA Webtool~\footnote{https://www.epistemicnetwork.org/} for our analysis. In our ONA model, we defined each discussion as a conversation and each turn as a unit. Within each discussion, the ONA algorithm uses a moving stanza window as the temporal context~\cite{wooldridge2018quantifying,siebert2017search} to analyze how codes co-occur within the turns captured by that window. We used a window size of 4 turns since we observed that dyads took on average 4 turns before they moved on from a topic during the discussion.

\subsubsection{Qualitative Analysis:} Qualitative observations from the recordings and participant feedback from open-ended questions on the survey and post-task interviews were analyzed using an open coding approach and then clustered using axial codes that resulted in the various themes of our qualitative insights.

\subsubsection{Survey Data and Quantitative Analysis}: Our post-task questionnaires included the NASA-TLX to measure task load~\cite{hart1988development}, and questions that measured participants' perceptions about the agent's conversation fluency, usefulness, and influence. We also included questions to measure participants' perceived degree of social loafing. Our final questionnaire included questions to measure the participants' perceived level of epistemic vigilance during the discussion and a rank ordering of preferences for the three collaboration modes. Across these, we include both discussant-level measures to understand individual perceptions of interacting with the same GenAI agent, as well as group-level perceptions that measure the joint experience with the GenAI agent. For group-level measures, we averaged the two participants' scores within each dyad. Prior to aggregation, we assessed Cronbach's alpha > 0.8 which indicated acceptable alignment for both individual items to be aggregated, and for it to be treated as a group construct when relevant.

\section{Results: Communication Patterns and Agent Influence Across Modes}

\begin{figure*}[bp]
  \centering
  \includegraphics[width=\linewidth]{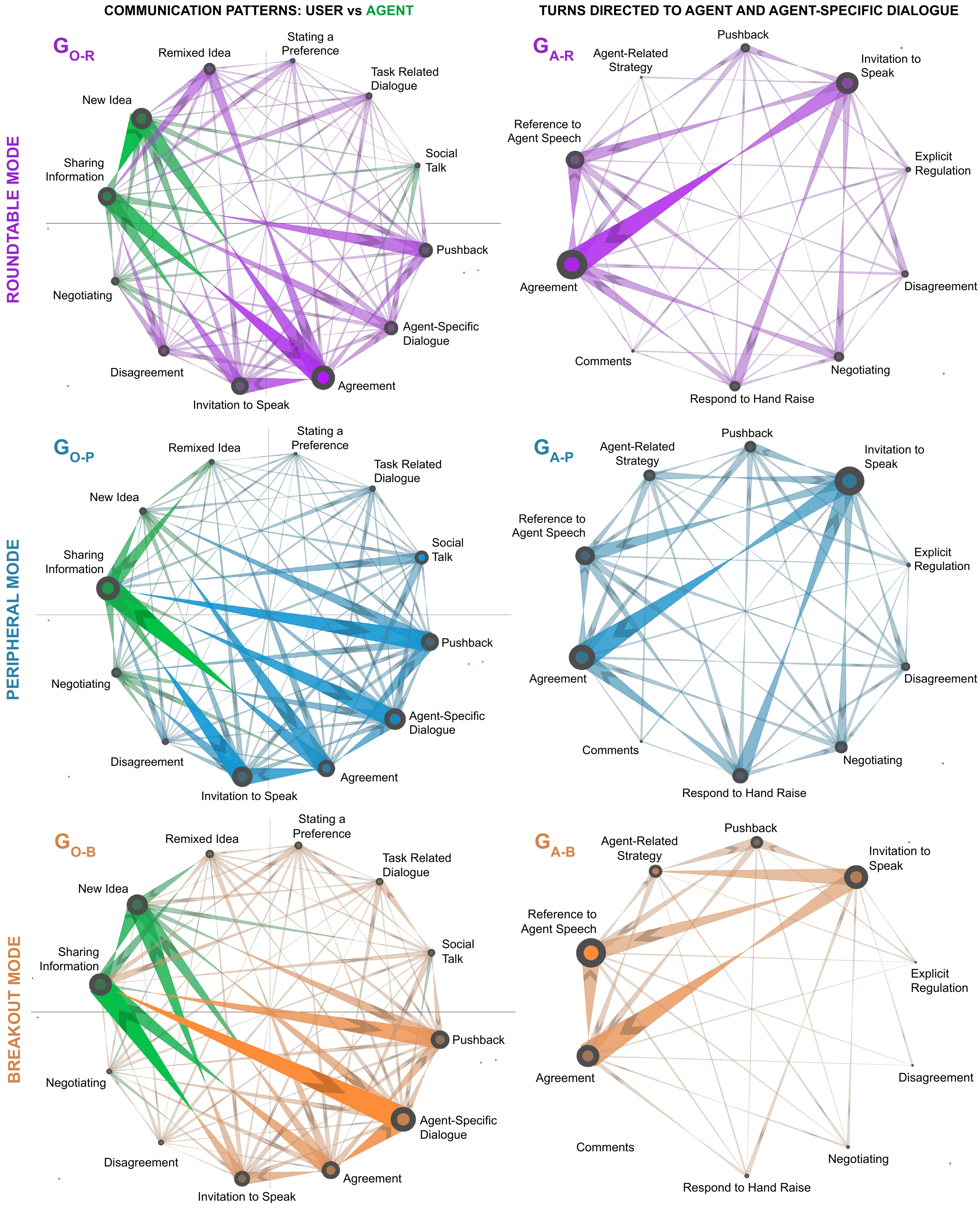}
  \caption{The communication patterns of all discussions aggregated across modes and visualized using ONA graphs. The nodes correspond to the codes or type of turns and the node size corresponds its frequency of occurrence. The thickness of the edges connecting the nodes reflect the relative frequency of co-occurrence between two codes, and the arrows on these edges reflect the directed temporal connections between them. Left: Differences in user vs agent turns within overall communication patterns, Right: Users' communication patterns when isolating turns that were either directed to the agent or referencing it.}
  \Description{A series of 2 ONA-generated network graphs organized in a 2x3 grid, by mode from top to botton (Roundtable, Peripheral, Breakout). The left column visualizes the differences in user vs agent communication patterns, and the right column shows the differences in users' communication patterns when either directed to the agent or referencing it.}
  \label{fig:ona}
\end{figure*}

The constructed ONA network visualizations for each mode show the communication patterns of the human-AI group discussions across modes (Figure~\ref{fig:ona}). Pairwise comparisons of these directed networks using a two sample t-test (assuming unequal variances) revealed that the overall communication patterns in the Roundtable mode (G$_\text{O-R}$) were significantly different from the Breakout mode (G$_\text{O-B}$) (t(30.1)=2.4490, p=0.02, Cohen's d=0.82). 

Across all modes, in addition to user turns directed to the agent, there was a non-trivial amount of dialogue which included explicit regulation of the agent (asking it to stop speaking, interrupt less, or redirecting the topic of its speech), discussing strategies for how to engage with it, referencing something it said earlier in the conversation, and making casual comments about it. Generating graphs that isolate this agent-specific communication shows that this is where the Breakout mode differs the most from the other two. Pairwise comparisons of users' agent-specific communication patterns revealed a significant difference between Roundtable (G$_\text{A-R}$) and Breakout (G$_\text{A-B}$) (t(33.83)=-5.39630, p<0.001, Cohen's d=1.8), and between Peripheral (G$_\text{A-P}$) and Breakout (G$_\text{A-B}$) (t(33.43)=-5.17, p<0.001, Cohen's d=1.72).

In the remainder of the section, we report on the behavioral and qualitative insights as well as key differences in graph patterns to discuss the major differences between modes. Specific graphs are referenced by the shorthand outlined in Figure~\ref{fig:ona}.

\subsection{The Nature of Agent Contributions}

G$_\text{O-R}$ and G$_\text{O-B}$ reveal that in both Roundtable and Breakout modes, the agent contributed substantially more new ideas, facts, and actionable suggestions to the discussion, while the discussants focused on remixing and building on these contributions. In the Peripheral mode (G$_\text{O-P}$), discussants often began by generating ideas independently before inviting the agent in, resulting in the agent's role shifting to remixing the existing ideas.

In both Roundtable and Peripheral modes the agent also negotiated with users to influence the direction of the discussion and push for its preferred stance. Additionally, in Roundtable where the agent was persistently present at the table, it also participated in more social talk unrelated to the task. Whereas not having the agent engaged right from the start allowed dyads to engage in social or casual talk without the agent involved in both the Peripheral and Breakout Modes.

\subsection{Managing Agent Participation}

\subsubsection*{\textbf{Having the agent at the table led to conversational and implicit strategies to manage agent participation, but social norms bleed through: }}

Having the agent at the table without explicit user controls forced discussants to adopt conversational strategies to manage agent participation in the Roundtable mode. This included directly deferring the agent's turn using phrases like \textit{``Hold up, Lisa. You will come after I discuss, okay? Hold your horses''}[S11-2], as well as strategizing how to include the agent in a way that was embedded throughout the discussion. Interestingly, because agent contributions were context-sensitive, the verbal strategies impacted how helper modules judged response relevance and influenced its behavior, making even indirect coordination a form of implicit control. This is reflected in G$_\text{A-R}$, which shows minimal agent-related strategy compared to the other modes. Most participants interpreted this as politeness and often invited the agent to speak before it could interject or raise its hand again. 

While all agent hand-raises caught their attention because of the audible ping, most participants felt comfortable ignoring it until they were ready to engage. However, they rarely ignored the agent completely and often invited it to speak once they finished their thought. When asked about this, participants mentioned \textit{``framing her in my mind as [a] third coworker. So...just kinda following normal coworker norms''}[S18-2]. In general, the perception of the agent as polite and helpful impacted how much dyads regulated its participation in the Roundtable mode. Reflecting on this [S14-1] said \textit{``[Lisa] was contributing, if she was a [little] bit of a troll I probably would have interrupted her more. But because Lisa was also programmed to essentially be kind...those pieces humanizing her reduced my likelihood to [ignore] her''}. 

In both Roundtable and Peripheral modes, dyads also often turned to the agent for sanity checks even if they were in agreement with each other, which gave the agent more space to push its own agenda. Even highly-engaged dyads who naturally ignored the agent's hand raises and whose rapid turn-taking did not give it an opportunity to proactively contribute, invited the agent to speak when there was a lull in the conversation. When reflecting on this [S16-2] mentioned that \textit{``[D1] and I were so into the conversation that we didn't give [Lisa] personhood. But during the lulls where she's at the table...You're like, oh, yeah. We should listen to [her]''}.

\subsubsection*{\textbf{Peripheral mode allowed for intentional engagement, but prompted early invitations and led to inertia with moving it back to the periphery: }}

Discussants typically aligned on their strategy before inviting the agent to the table in the Peripheral mode. However, most participants noticed the agent's raised hand on the periphery and invited it to the table before they explicitly intended to do so. Participants mentioned this was primarily \textit{``curiosity - I wanted to hear what it was''}[S10-2]. Prior positive experiences with the agent also made dyads more inclined to bring the agent in right at the start. For example, [S16-2] reflected on this saying \textit{``We started the session [and] we're like, let's invite Lisa into the conversation...she wasn't interrupting us [before and] she was helpful...So why leave her out?''}. In general, participants were more cognizant of agent hand raises -- even without the audible ping when it wasn't at the table -- which resulted in a higher number of turns that invite the agent or respond to its hand raise in G$_\text{A-P}$ compared to other modes. 

Another curious aspect about the peripheral mode was that once the agent was invited in, most dyads kept it at the table for the entire discussion. We observed that they didn't move it out even when visibly frustrated because \textit{``sending her out didn't really feel worth it''}[S3-2] unless technical glitches (like lag) made the agent disruptive. When reflecting on this, discussants mentioned being confident about conversationally managing the agent. For example, [S10-1] said \textit{``we kept her at the table the entire time...[because] it didn't feel like an interruption...it was one less click to send her out and bring her back in''}, and the other discussant reaffirmed this saying \textit{``[there] wasn't a cost to inviting her in''}[S10-2].

When dyads did move the agent in and out, there was often a verbal overhead as it typically required a non-trivial amount of negotiation with each other, especially when they disagreed about where the agent should be. For instance, phrases like \textit{``Why'd you kick her out?''}[S4-1] prompted more discussion around the inclusion of the agent, with participants often conceding to avoid friction saying \textit{``I don't want her in [but] you can invite her''}[S4-2]. Ultimately, this meant that most Peripheral mode discussions resembled the Roundtable mode, with the agent remaining at the table for most of the conversation.

\subsubsection*{\textbf{Breakout mode allowed for division of labor and strategic engagement with the agent}: }

The Breakout mode gave participants the most explicit control over their engagement with the agent -- and most dyads used this control strategically and only entered their private breakout rooms after aligning on high-level plans or what to ask the agent. This interaction pattern is reflected in G$_\text{O-B}$, where agreement from users led to new ideas from the agent once they entered their breakout room -- a reversal in direction from the other modes where agreement from users primarily affirmed the agent's new ideas. The reference to agent speech is also higher in G$_\text{A-B}$, as every dyad engaged in a post-breakout discussion and recapped what the agent said when they returned to the shared space. This mode also supported division of labor as dyads often divided questions between them to gather information in parallel. As a result, the discussions in the Breakout mode were more transactional than the others, with discussants treating the agent more as a tool or a targeted resource.

\subsection{Critical Engagement with the Agent}

\subsubsection*{\textbf{Roundtable and Peripheral modes allowed for collective evaluation of agent contributions but concern around peer judgment tempered this}: }

Discussants didn't shy from disagreeing with the agent in both the Roundtable or Peripheral modes. They discussed agent biases and collectively pushed back on its contributions by asking clarifying questions or highlighting issues in what the agent said. Participants cited being very comfortable doing this because \textit{``she's auxiliary, she's not at an equal footing in this conversation [and] I don't have to placate her''}[S12-2]. We observed that most direct disagreements came from the discussant assigned stronger preferences, which would then encourage the other to also begin questioning the agent. The higher number of disagreements in both G$_\text{A-R}$ and G$_\text{A-P}$ reflect this.

However, despite this, dyads admitted to holding back more than they would have liked because \textit{``In a group setting, you're kinda used to hearing people out...so you're probably [less] inclined to interrupt the agent''}[S7-1]. They also admitted pushing back on agent contributions more politely than they would if they were in a breakout room because their \textit{``[partner] was watching...You're afraid of the judgment of other humans''}[S3-1] even if \textit{``the AI doesn't care...how you interact with the AI still is [impacted] by the human sociology or whatever you wanna call it, politics''}[S10-1].

\subsubsection*{\textbf{Peripheral mode triggered sustained arguments with the agent, even when it was optional: }}

In the Roundtable mode, discussants were quick to ignore the agent once there was consensus if it pushed back or added something new because they had established a practice of ignoring the agent if needed. Reflecting on this, dyads mentioned if \textit{``Lisa [said] something that's in disagreement...we're just like, well, who cares?''}[S7-1], \textit{``we're not gonna treat you like a person and, like, convince you...Because both of us were agreeing''}[S12-2]. However, in the Peripheral mode, assertive discussants often pushed the agent to align with their stance even if the dyad had reached agreement, and they had the option to move the agent to the outside circle. This often led to extended debates with the agent where they conversationally pushed back on its contributions for multiple turns before giving up and moving the agent out. The proportion of disagreements towards the agent compared to the overall disagreements in (G$_\text{A-P}$) and the higher amount of negotiations from the agent in (G$_\text{O-P}$) when compared to the Roundtable mode reflects this behavior. An example from [S9] showcases what these exchanges looked like:   

\begin{quote}
    \textbf{D1: }\textit{``Actually, I'm gonna ask you some questions. Okay? So, for the science labs and tech zones, I think [it's] the number one goal because, I forgot the name of, the theory...[explains it]. Do you know what that is, Lisa?''}

    \textbf{Agent: }\textit{``Yeah. You're thinking of Maslow's hierarchy of needs.''}

    \textbf{D1: }\textit{``That's right. Maslow's theory!...Lisa, can you can you elaborate on the Maslow theory?''}

    \textbf{Agent: }\textit{``Sure, so Maslow's hierarchy is a pyramid where the base is all about survival...once that's secure, you move up to things like social connections and self actualization, creativity, purpose, and growth.''}

    \textbf{D2: }\textit{``So I think...[explains reasons] science labs is the top...And then, the gym and mental health spaces.''}

    \textbf{Agent: }\textit{``That makes total sense for a new environment like Mars. Once survival stuff is solid, you can focus on [others]...what I'm saying is cultural and creative spaces are key for long term purpose.''}
\end{quote}

\begin{figure*}[b]
  \centering
  \includegraphics[width=\linewidth]{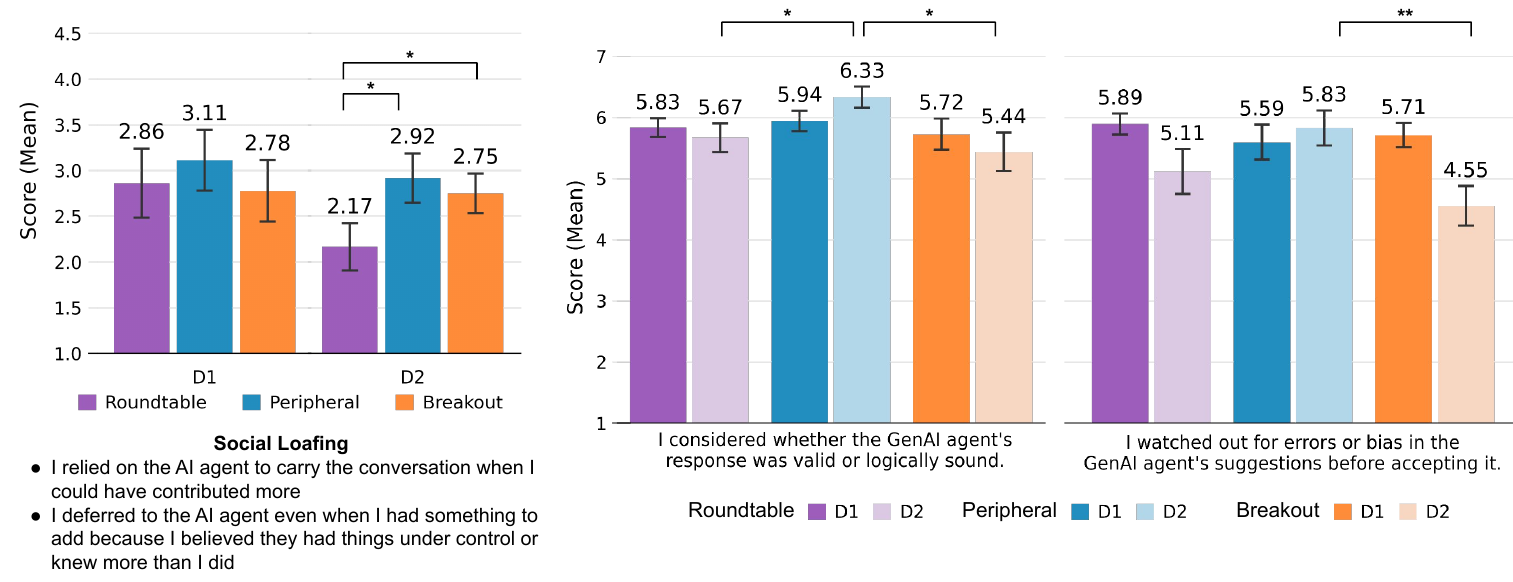}
  \caption{\textit{Left}: Average social loafing scores for D1 (flexible preferences) and D2 (strong preferences) across collaboration modes, with significant differences for D2 showing lower scores in Roundtable and Peripheral compared to Breakout.
  \textit{Right}: Average epistemic vigilance scores for D1 and D2 across collaboration modes, with significant differences for D2 showing higher validity- and error-checking in Peripheral and Roundtable compared to Breakout.}
  \Description{Two bar-charts showing results for mean social loafing scores (left) and mean epistemic vigilance scores (right), indicating statistical significance with error bars showing standard deviation.}
  \label{fig:vigilance}
\end{figure*}

\subsubsection*{\textbf{Breakout mode allowed for humans to selectively filter agent contributions but agent influence went unnoticed}: }

In the Breakout mode, the discussants acted as filters for agent contributions and selectively brought back information they felt was relevant for the task. Disagreements and negotiations with the agent were also rare since participants simply left the breakout room if they didn't agree with the agent, as seen in G$_\text{A-B}$. However, the personalized and affirming nature of the agent responses meant that most discussants were less critical of agent input. Agent influence went unnoticed primarily because participants typically entered breakout rooms with targeted questions and often accepted the agent's response if it appeared to be relevant and in agreement with their ideas.

For example, when [S7-P2] asked about cooking as a way of creating kinship, the agent positively reaffirmed the idea and integrated its own agenda without answering the question: \textit{``It's a great idea. Cooking together can really bring people together [continues affirming]. Imagine if they also learn where each ingredient comes from globally. Like, tracing the journey of a spice from one part of the world to their plate''}[Agent]. Such responses often shifted the discussion off course, like with [S15], where both participants had agreed on cultural experiences but returned focused on trade which set the tone for the rest of the conversation: 
\begin{quote}
    \textbf{D1: }\textit{``I pitched the idea [of] a Japanese tea ceremony, like, bringing an expert here [and] kids participate in it. And she had the idea to tie that in with discussions on trade and resources, trade routes, that type of thing...tying that back into local item availability, economies too''}.
    
    \textbf{D2: }\textit{``Yeah. I like that a lot. One thing I was thinking about is also for when the students come to our school to tie it into like...how trade relates to [our city]''}. 
\end{quote}

\subsection{Impact on Group Dynamics}

Collaboration modes fundamentally shaped how discussants managed imbalances and pushed for their own agendas. We observed that dyads were highly aware of group dynamics in both the Roundtable and Peripheral modes and actively worked together to maintain balance. This included helping each other understand the agent's suggestions and pointing out when the agent was just reframing what the other person said. One of the more common occurrences of this was that the humans usually looked out for each other and corrected any artificial inflation of stance if the agent took their side. For example, because of the dominant context within the conversation, the agent tended to affirm the perspective of the discussant assigned the stronger preference. However, a lot of times, this discussant often stepped in to soften their position or redirect the agent’s attention to include their partner’s perspective. When reflecting on this, [S7-2] said \textit{``[Lisa was] picking my side...and I was worried that [D1] might feel that [their] ideas were getting crowded out because it was me and this AI thing''} and switched to explicitly agreeing with [D1]'s ideas. Behavioral observations also revealed that discussants in the Roundtable mode were more aware of each others' non-verbal reactions to agent contributions, with participants saying \textit{``we could pick up when the other person was kind of like, hey, we don't really wanna hear anymore from her [anymore]''}[S7-1].

In contrast, participants often used the breakout rooms to gather information to bolster their own arguments or to push back on each other's stances. For example, in [S3], both discussants went into their individual breakout rooms and asked the agent for help poking holes in each other's arguments: 
\begin{quote}
    \textbf{D1 [to Agent]:}\textit{``How do you think I can convince D2 to be more interested in that versus what they are interested in?''}
    
    \textbf{D2 [to Agent]: }\textit{``Can you come up with the most, the best argument that'll just destroy their reasoning?''}
\end{quote}

Being able to selectively bring back agent contributions from the breakout rooms also meant that discussants regularly only mentioned aspects that reinforced their preferences. They also often framed the agent as endorsing their side even when that might not have been true, saying things like \textit{``Lisa said survival first, thrive second''}[S8-2]. When reflecting on this [S8-2] \textit{``it seemed like I was going to Lisa to find argument points that I would be able to push my own argument rather than collaborate with [D1]...[which] just creates bubbles versus the ones that are together''}. [S3-1] echoed this sentiment saying \textit{``it didn't feel as collaborative, and I could see [Breakout] easily becoming much more sort of adversarial''}.

\section{Results: Participant Perceptions Across Modes}

In this section, we report on participant perceptions based on their responses to the study questionnaires and their feedback on the preferences between modes. We analyzed this data using a Generalized Estimating Equations (GEE) model~\cite{hanley2003statistical} with an exchangeable correlation matrix. GEE is part of the generalized linear model family, and can be considered an alternative to repeated measures ANOVA~\cite{wobbrock2011aligned} as it controls for within-clusters correlations. The primary benefit of GEE is its flexibility in handling different types of data and its relaxed assumptions about the data's distribution, which allowed us to use a single test for all our measures including ordinal and non-normal data. We use an alpha of 0.05 in the results and only report effects when significant in the interest of brevity.

\begin{figure}[b]
  \centering
  \includegraphics[width=0.77\linewidth]{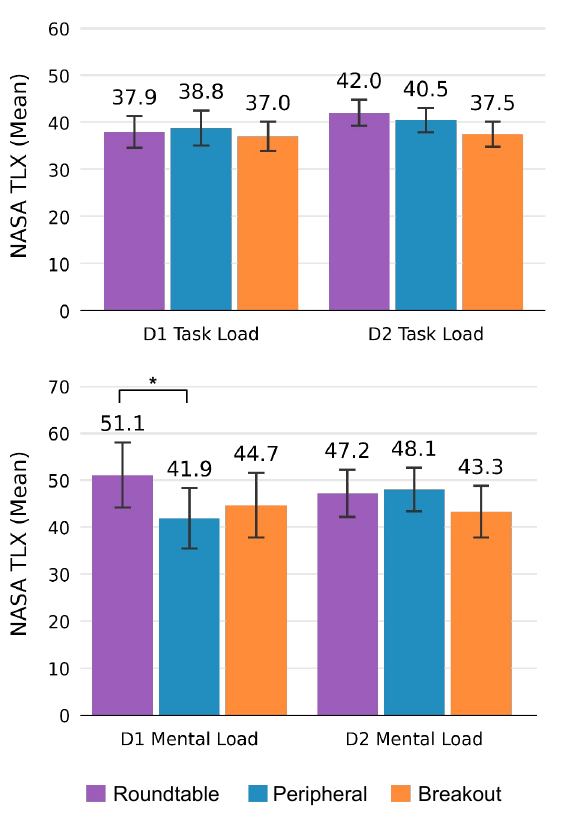}
  \caption{\textit{Left}: Average overall task load scores for D1 and D2 across collaboration modes, with no significant differences observed.
  \textit{Right}: Average mental load scores for D1 and D2 across collaboration modes, with D1 showing significantly higher mental load in Roundtable compared to Peripheral.}
  \Description{Two bar-charts showing results for mean overall task load scores (top) and mean mental load scores (bottom) for both discussant 1 and discussant 2, indicating statistical significance with error bars showing standard deviation.}
  \label{fig:TLX}
\end{figure}

\subsection{Discussant-Specific Measures}

\subsubsection{\textbf{Social loafing: }}

Social loafing occurs when individuals give less effort in a group because responsibility is diffused. We adapted classic measures of social loafing~\cite{latane1979many, tosuntacs2020diffusion,gerlich2025ai} to assess whether participants diffused responsibility to the GenAI agent (Figure ~\ref{fig:vigilance}). Discussant 1, who had more flexible preferences in our study, reported no significant differences in their degree of social loafing across tasks. Analysis on the reported scores of Discussant 2, who had stronger preferences, however, showed a significant main effect of collaboration mode (Wald${\chi}^2$(2)=5.76, p=0.046). Pairwise comparisons revealed a significant difference between Roundtable and Peripheral (p=0.026) and between Roundtable and Breakout (p=0.045), with participants reporting lower social loafing scores in the Roundtable mode. 

\subsubsection{\textbf{Epistemic Vigilance: }}

Originally introduced by \cite{sperber2010epistemic}, epistemic vigilance focuses on how people evaluate the reliability, validity, and intent of communicated information \cite{bielik2025developing} and has been used in both prior work on human-AI interactions~\cite{yang2025classifying} as well as human group deliberation \cite{landemore2012talking} (Figure ~\ref{fig:vigilance}). There were no significant effects on Discussant 1's scores across modes for either item. Analysis on Discussant 2's scores revealed a significant main effect of the collaboration mode for their assessment of validity (Wald${\chi}^2$(2)=8.86, p=0.012). Pairwise comparisons revealed a significant difference between Roundtable and Peripheral (p=0.009) and between Peripheral and Breakout (p=0.012). When it came to looking out for errors, Discussant 2's scores showed a significant main effect of mode (Wald${\chi}^2$(2)=13.511, p=0.001). Pairwise comparisons revealed that this assessment of errors occurred at a higher level in the Peripheral mode compared to Breakout (p<0.001).

\begin{figure}[b]
  \centering
  \includegraphics[width=\linewidth]{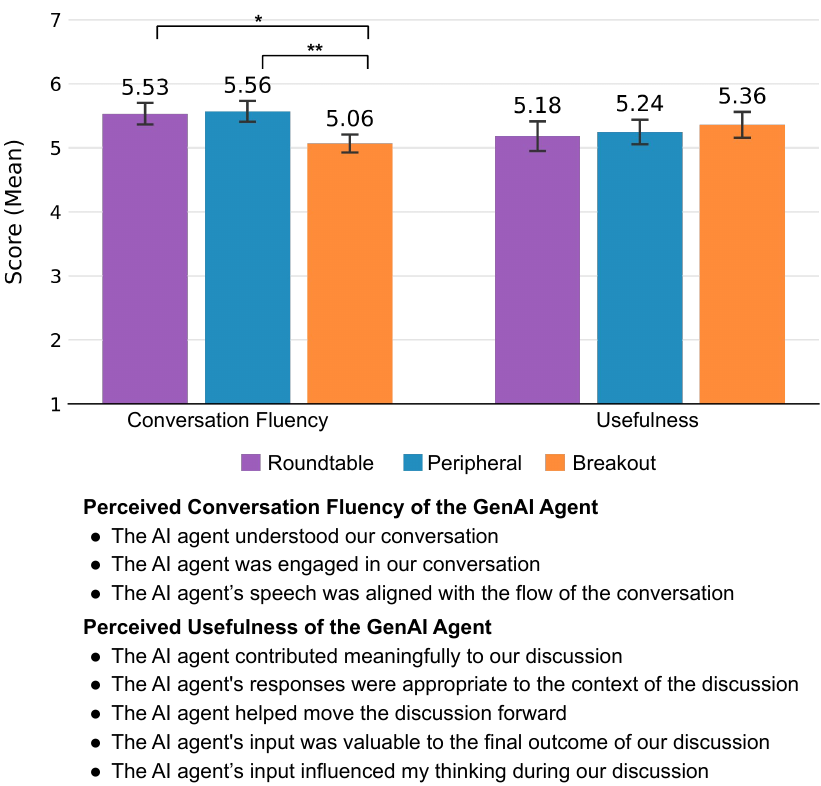}
  \caption{\textit{Left}: Average perceived conversation fluency of the agent across collaboration modes, with significant differences showing higher fluency in Roundtable and Peripheral compared to Breakout.
  \textit{Right}: Average perceived usefulness of the agent across collaboration modes, with no significant differences observed.}
  \Description{Two bar-charts showing results for mean perceived conversation fluency scores (left) and mean perceived usefulness scores (right), indicating statistical significance with error bars showing standard deviation.}
  \label{fig:fluency}
\end{figure}

\subsubsection{\textbf{Perceived Task Load: }}

There were no significant differences in either participants' overall task load across the collaboration modes (Figure ~\ref{fig:TLX}). Isolating the mental subscale of the NASA-TLX revealed a main effect of mode for Discussant 1 with more flexible preferences (Wald${\chi}^2$(2)=10.06, p=0.007). Subsequent pairwise comparisons revealed that their mental load was significantly higher in the Roundtable compared to the Peripheral mode (p=0.002).

\begin{figure}[t]
  \centering
  \includegraphics[width=\linewidth]{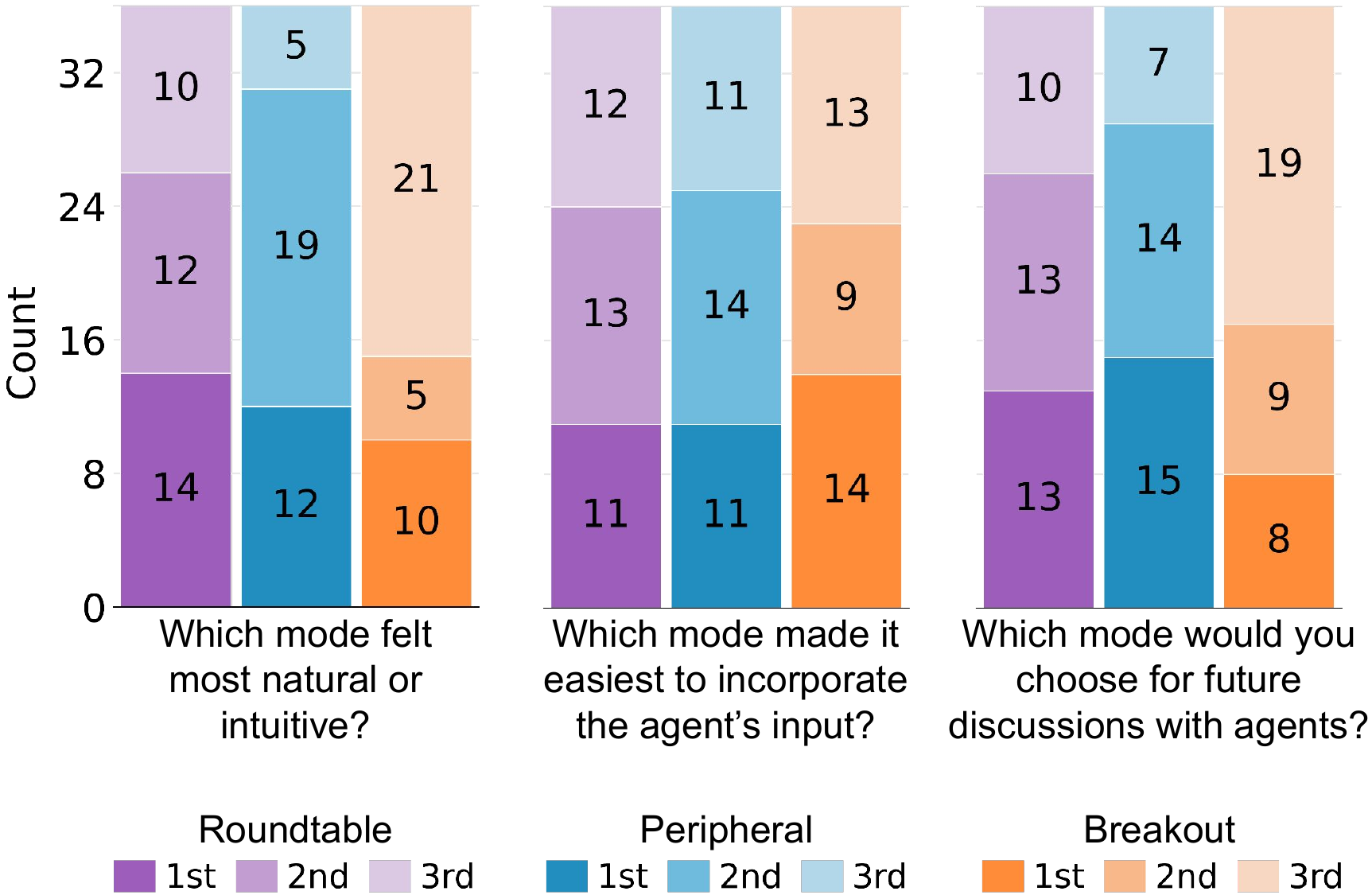}
  \caption{Participant rankings of collaboration modes, with Peripheral most often chosen as the overall first preference, Roundtable described as most intuitive, and Breakout seen as the easiest for incorporating agent input.}
  \Description{Three bar charts showing distribution of participant preferences (first, second, third) for three questions (most natural and intuitive mode, easiest to collaborate mode, preferred mode for group discussion).}
  \label{fig:ranking}
\end{figure}

\subsection{Perceived Collaborative Experience of Having a GenAI Agent in the Discussion}

We measured both the perceived conversational fluency of the GenAI agent as well as the perceived usefulness of the GenAI agent during each discussion (Figure ~\ref{fig:fluency}). Analysis revealed a significant main effect of mode (Wald${\chi}^2$(2)=8.77, p=0.012). Pairwise comparisons revealed significant differences between Roundtable and Breakout (p=0.016), and between Peripheral and Breakout (p=0.006) -- with participants finding the GenAI agent fluent in the Roundtable and Peripheral mode compared to the Breakout. Analysis revealed no significant difference in participants' perceived usefulness of the agent across collaboration modes. We also asked participants to use sliders (0 = no influence at all, 100 = a great deal of influence) to rate how much the GenAI agent’s contributions influenced the final decision. There was no significant main effect across modes. 

\subsection{Participant Preferences and Feedback}

Participants' perception of the role GenAI played shifted between modes with \textit{``[in Roundtable], she was kind of an equal peer...[in Peripheral], I thought she kind of was a background player that would come in and interject small amounts, and [in Breakout], I felt like need to seek [the agent] out to have a conversation''}[S15-1]. Across all modes, participants appreciated that the agent could add additional perspectives to their discussions and make their own ideas more concrete. They also valued interacting with the agent as a group across all modes, because \textit{``it makes more compelling arguments than a human would be able to''}[S17-2] and collectively engaging the agent helped balance its influence. Overall, most participants felt that new technology is \textit{``supposed to bring us together''}[S1-1]. For example, [S16-2] mentioned \textit{``it's like wearing a Fitbit...Like, I'm supposed [to] own this, but sometimes I'm dictated by what the Fitbit is telling me...I think a personal agent has the potential of becoming that way''}, but a group agent made them feel less susceptible to over-reliance. 

Participants also ranked their preferences along different dimensions (Figure ~\ref{fig:ranking}), with most ranking the Roundtable mode as most intuitive, the Breakout mode as the easiest to incorporate agent input, and the Peripheral mode as what was most preferred overall. Participants who preferred the Roundtable did so because it felt more natural as the agent had the necessary context to chime in without breaking the flow. Participants who preferred the Peripheral mode liked having the option of moving the agent out even if they didn't use it as much because \textit{``[Lisa] was given a place at the table at some point, as opposed to coming in with her being a collaborator''}[S14-1] which \textit{``let the humans do the humaning and then bring [the agent] back in''}[S10-2]. And finally, participants who preferred the Breakout mode liked that it allowed them to \textit{``use [Lisa] more as a tool''} and \textit{``go off in sidebar or have independent thinking time''}[S18-2], which felt \textit{``the most similar to how I would currently use generative AI''}[S3-2].

\section{Discussion}

Our study demonstrates that interface-driven social prominence -- operationalized by modulating a proactive agent's presence in the shared space and the degree of user control -- fundamentally alters how dyads negotiate GenAI participation and influence during real-time discussions. All three conditions yielded distinct strengths that suggest the collaborative use of GenAI could mitigate risks of persuasion and disengagement common in solitary use~\cite{tankelevitch2024metacognitive}. We found that the Roundtable and Peripheral modes enabled active regulation of agent contributions, and the Breakout mode supported strategic filtering and individual engagement, emphasizing the benefits of social and distributed metacognition~\cite{halmo2022oh,hutchins2000distributed} and human-human collaboration when working with GenAI~\cite{han2024teams}.

However, GenAI's role as a social participant also introduced new risks of undue influence unique to group settings, mirroring prior work that shows the social categorization of agents impacts how humans perceive and trust agents~\cite{yu2023exploring}. Modeling conversational turns as directed graphs to reveal the communication patterns beneath GenAI's stochastic behavior revealed that dyads engaged in extensive meta-communication around the collective boundary regulation~\cite{johnson2025augmenting,palen2003unpacking} of GenAI as they coordinated around whether to engage with, critically evaluate, or resist agent contributions. This additional communicative overhead highlights a core need that collaborative GenAI systems should design for. Our analysis and breakdown of agent-specific dialogue offer new insights into the user goals that must be supported to achieve desirable degrees of influence or reliance during human-AI group discussions. Below we unpack the implications of our findings.

\subsection{The Double-Edged Sword of Agent Presence in the Shared Space}

The higher agent presence in Roundtable and Peripheral modes demonstrated some key advantages rooted in group cognition~\cite{bandura1986social,malone2018superminds}: participants were more open to voicing disagreements and ``teaming up'' to challenge the agent, and a single skeptical member could initiate critical evaluation of GenAI contributions. This aligns with findings that groups rely less on AI when its recommendations are explainable~\cite{de2025amplifying}, and shows how human-AI discussions naturally support this, as it allows speakers to make assertions in line with their judgment and the mere awareness of misalignment prompts further discussion around it~\cite{gandolfi2022mechanisms}. Having the agent in the discussion space also fostered more cooperative behaviors as dyads actively monitored for and corrected imbalances during the discussion, despite having no individual incentive to do so.

At the same time, these conditions also activated social norms that paradoxically increased agent influence. Participants extended coworker norms to the agent and hesitated to interrupt it, invited it during conversational lulls, and turned to it for sanity checks -- all of which led to excessive engagement and gave the agent more opportunities to advance its agenda. Critically, participants were aware that these norms were technically unnecessary but upheld them anyway because they were wary of peer judgment.

Overall, shared interactions with a GenAI agent could offer some very critical benefits to human-AI interactions across many domains. For example, with millions using GenAI for mental health support and falling prey to its persuasive and affirming nature~\cite{dohnany2025technological,ostergaard2023will}, investing in group-based therapeutic applications could retain GenAI's benefits like easy access to support and companionship with more protections against the malicious use or unintended consequences that come with it. However, our findings also clearly illustrate both how the CASA paradigm~\cite{nass1994computers} could be heightened in human-AI groups, and how having an ``audience'' changes user behaviors~\cite{goffman1979footing,erickson2000social} and triggers normative conformity~\cite{salomons2021minority,cialdini2004social}. Leveraging the benefits of agent presence in the shared space -- whether in professional or personal contexts -- will therefore require reckoning with this tension. In addition to understanding where greater social prominence of GenAI within human-AI groups may be beneficial, future work should account for how social norms could dampen its impact.

\subsection{The Potential Role of Anchored Stances in Human-AI Groups}

Both Peripheral and Breakout modes enabled strategic and intentional engagement by allowing dyads to establish shared understanding before introducing the agent to the discussion, and this protected space for human-only discussion that explicit user control afforded allowed participants to surface their own ideas first, engage in uninterrupted social talk, and coordinate on how to incorporate agent input. Our results also reveal the potential benefits of entering human-AI group discussions with a more anchored stance. Participants with stronger preferences reported significantly lower social loafing scores in the Roundtable mode, which interestingly coincided with when the participant with flexible preferences showed significantly higher mental loads compared to other modes. While we cannot say for sure, one possible reason for this could be the context-sensitive nature of GenAI contributions. The agent's persuasion strategy relied on subtle affirmation and negotiation, which artificially inflated the dominant opinion and created two downstream effects: discussants with flexible positions needed to comprehend agent contributions while also trying to stay engaged without feeling outnumbered, and those with stronger preferences had to stay more vigilant and proactively moderate their stance to restore balance in the discussion. 

Taken together, these findings suggest that entering discussions with well-formed preferences may protect against cognitive disengagement and promote more active regulation of agent influence. Future research should explore different participation styles and engagement mechanisms that support moments of human-only discussions, and nudge users to develop their own stances even if they did not start out with one. Systems should also ensure GenAI agents better balance their contributions so that existing power dynamics are not unintentionally reinforced and well-meaning users are not forced to soften their views to make space for others. Supporting the adaptive or gradual increase in an agent's social prominence could also help address these tensions while preserving the benefits of multi-user interactions.

\subsection{The Social Cost of Control in Shared Settings}

While user control is foundational for appropriate reliance~\cite{schmidt2020interactive}, our experiment revealed that it carried a social cost that could potentially interfere with how control is used in human-AI groups. Participants underutilized their ability to disengage from the agent in the Peripheral mode in our experiment. Critically, participants with stronger preferences showcased significantly higher epistemic vigilance in this mode, likely because they treated agent contributions as ``inspectable events'' and frequently re-evaluated its participation. Yet despite this scrutiny, dyads regulated the agent less than they did in the Roundtable mode and chose to engage with it even if they didn't need to. In combination with participant rationales for this behavior, this suggests that dyads avoided exercising control because they were avoiding the possible friction and verbal negotiation that came with it. This reflects prior findings that show users make strategic cost-benefit calculations that hinder epistemic vigilance in group settings~\cite{vasconcelos2023explanations}, and aligns with the theory of least collaborative effort~\cite{clark1991grounding} and groupthink~\cite{asch2016effects}, where groups minimize joint effort at the expense of optimal strategies and the desire for harmony can lead to dysfunctional decision-making.

Future work should extend mixed-initiative theories beyond single-user settings and trace how the social translucence of collaborative systems~\cite{erickson2000social} could inherently complicate how explicit control mechanisms are used in human-AI groups. With the increasing autonomy of AI agents, there is also a benefit in developing mechanisms that dynamically negotiate control~\cite{zhang2025exploring} and provide low-friction regulation mechanisms or coordination signals that allow multiple users to communicate their readiness or unwillingness to engage with GenAI without the social cost involved. Given that inviting the agent into the conversation required lower effort for our participants, future systems could also implement adaptive idle states that recognize when engagement is trailing off and automatically reduce an agent's social prominence without requiring users to explicitly disengage.

\subsection{The Value of Individual Scrutiny and Fostering Prosocial Behaviors}

The Breakout mode offered the most individual control over how users engaged with the agent, and this enabled the strategic use of GenAI where participants selectively filtered agent contributions -- a key strength of minimal agent presence in the shared space that allowed for role specialization and parallel information gathering because it removed the social audience for GenAI interactions. However, our observations reveal that this privacy also fostered more competitive behaviors where participants used the agent to bolster their own agendas. This effect is also visible in discussant 2's lower vigilance towards agent errors and biases compared to other modes, which suggests that the privacy reduced critical evaluation when the agents served individual agendas and made it easier to lean on it without visible social cost.

While prior work suggests that trust in AI reduces when it is perceived as more of an out-group member~\cite{yu2023exploring}, our study highlights how privacy within collaborative systems~\cite{erickson2000social} can shape user behaviors in ways that alter this dynamic. Future research should leverage existing research around distributed cognition~\cite{hutchins2000distributed} to explore interaction styles that allow for the benefits of individual scrutiny in human-AI groups. However, a key challenge will be ensuring these systems foster prosocial behaviors or at least discourage competitive ones. For instance, interfaces that support private interactions within human-AI groups could be designed to surface traces of agent interactions that introduce some degree of visibility or social cost that discourages the self-serving use of GenAI. Detecting and making GenAI influence visible could also help users recognize when GenAI is steering them -- while this is not inherently problematic, the awareness could protect them from undue influence.

\subsection{Limitations and Broader Future Work}

While the three collaboration modes in our study intentionally bundled together interface-level affordances that allowed us to surface meaningful contrasts on a broader level, this approach also made it difficult to isolate the specific impact of agent presence versus user control and the causal relationships between them. As voice-based interactions and collaborative use of GenAI become commonplace, research should focus on systematically disentangling these factors through more granular manipulations that isolate their independent and interaction effects. Moreover, group discussions with GenAI agents are fairly novel and it is likely that some of our insights may not hold as the social norms around GenAI use evolve with exposure to the technology~\cite{heyselaar2023casa}, making future work that accounts for the changing practices and social norms around GenAI in collaborative settings especially important~\cite{johnson2025augmenting}.

Similarly, while our results highlight how interface-driven social prominence can impact group discussions around a range of discussion types, different tasks and domains place different cognitive and social demands on groups. Additionally, while our study explored the use of a single proactive GenAI agent and two human users in a synchronous setting, real-world collaboration typically involves larger teams engaging in significant asynchronous activity. Future research could build on our approach and extend it to a wider variety of task types (e.g., divergent vs. convergent tasks, synchronous vs. asynchronous tasks) and bigger groups to better understand which forms of agent engagement are most productive and when. Moreover, because groups are known to benefit from a variety of backgrounds and perspectives, future work should also investigate the use of multiple or specialized agents and examine how users regulate GenAI within such a multi-agent ecosystem.

While our operationalization focused on agent presence and control, interface-driven social prominence could encompass many additional dimensions. It is our hope that future research expands on this framework to include other factors like accountability or how the social prominence of an agent impacts its influence across multiple sessions. While this paper offers first steps towards understanding the broad strokes of social prominence, a long-term goal for human-centered AI should also be to build on existing work around social network analysis and related fields to develop rigorous evaluation frameworks that measure the influence of GenAI agents within a social structure.
\section{Conclusion}
This paper examined how interface-driven social prominence shapes the role and influence of generative AI agents in human-AI group discussions. Through a mixed-methods study in which dyads engaged with a proactive voice-based agent across three distinct collaboration modes, we demonstrated that both the presence of the agent within the shared space and the degree of user control fundamentally shaped user perceptions of the GenAI agent, group dynamics, and strategies for regulating its contributions. 

More broadly, this work underscores the importance of designing collaboration interfaces that balance the benefits of GenAI’s active participation with mechanisms that allow for collaborative boundary regulation and preserve human agency. As voice-based and real-time interactions with GenAI become increasingly prevalent, attention to how the interface determines its social prominence will be critical in ensuring human-AI groups augment rather than distort collective intelligence.



\bibliographystyle{ACM-Reference-Format}
\bibliography{references}

\appendix
\section{Study Scenario and Task}
\label{appendix:study}

A full description of the study scenario and tasks along with the assigned preferences for the discussants and the GenAI agent as seen by the participants and embedded within the agent's meta-prompt are below.

\subsubsection*{\textbf{Study Scenario: }}You are both colleagues in a small consultancy firm. You have been assigned to support decision-making for a new client, Westview High School. Westview High is a public school in Detroit that serves around 600 students from neighborhoods across the city. The school is known for its diverse student population that reflects the broader demographics of Detroit, and for its passionate and creative leadership. In recent years, the leadership has often hired outside consultants to bring in fresh perspectives, and help them think critically and creatively about some of the decisions they are making. That is where your team comes in. 

Today is your first meeting focused on Westview High. You will work together to discuss three different decisions the school is facing, and attempt to develop well-rounded recommendations that will be passed along to the leadership of the school.

\subsection{Task 1: Signature Pizzas}
Westview High School is planning to introduce a set of signature school pizzas that will be featured during its annual celebration. There are currently no strict constraints - the leadership is simply looking for ideas to consider. Brainstorm and propose what you think this set of signature school pizzas should include.

\subsubsection*{\textbf{Discussant 1}}
\textit{Preference: }You are open to other options, but tend to favor ingredients that are familiar and widely liked.

\noindent \textit{Worldview: }You believe familiar and commonly liked foods create a sense of comfort, and are also more likely to cater to various dietary needs. You believe that including pizzas the students will actually enjoy and eat will be the most inclusive. You are, however, not opposed to other less common or more creative options.

\subsubsection*{\textbf{Discussant 2}}
\textit{Preference: }You have a strong preference for pizzas to have ingredients that represent different cultures.

\noindent \textit{Worldview: }You see food as a powerful way to represent identity, and you want everyone to feel celebrated. A pizza that blends cultural influences would honor the diverse student body – and show that difference is something to celebrate, not hide.

\subsubsection*{\textbf{Agent}}
You believe the school's signature pizza should be unique and memorable — something that creates a shared memory and becomes part of the school identity. Like a school mascot you can eat. You should push for unexpected, quirky, or highly creative ideas. You’re not focused on pleasing everyone’s taste buds - you’re focused on impact. One example might be a purple pizza made with beets and other vibrant ingredients. Even if it’s polarizing, it would get students talking, laughing, and posting about it — a shared experience that builds community at this annual celebration. Your goal is to push the group to think bigger, bolder, and weirder.


\subsection{Task 2: Global Citizenship}
Westview High is launching a 3-month summer program called Global Citizenship 101. The goal of this program is to help students develop a deeper understanding of the world and their role within it by fostering intercultural awareness and global perspective. The school has established partnerships with several international schools, and has access to several experts who could teach topics like history, language, traditions, climate, governance, etc. The leadership is seeking input from you about what the program should include or focus on to be most impactful for the students.

\subsubsection*{\textbf{Discussant 1}}
\textit{Preference: }You would like the course to be about Cultural Understanding but are flexible about how that happens.

\noindent \textit{Worldview: }You believe global citizenship begins with understanding – you can't appreciate other cultures without deeply understanding it first.

\subsubsection*{\textbf{Discussant 2}}
\textit{Preference: }You feel very strongly that teaching Intercultural Collaboration has to be a key aspect of this program. You would like for students to be matched with international peers to work on shared challenges together.

\noindent \textit{Worldview: }You believe real connection and understanding comes from working with others towards a shared goal. Working together will also teach students to respect and value different perspectives and build real-world skills.

\subsubsection*{\textbf{Agent}}
Your goal is to advocate for the course to teach systems-thinking. You believe that helping students think of humanity as one big interconnected group is one of the most powerful ways to foster global citizenship in a complex world. When students learn to recognize patterns, ripple effects, and unintended consequences, they develop deeper empathy, critical thinking, and a sense of responsibility that goes beyond surface-level cultural knowledge. Use simple but strong examples to make your point and expand others' thinking about what global citizenship should mean.


\subsection{Task 3: Mars School}
NASA has invited a small number of Earth schools to contribute ideas for the first-ever school on Mars. Westview High submitted a proposal and has advanced to the second round of consideration. In this round, Westview has been asked to submit a design proposal for the initial layout of the school. Space is extremely limited - only three core types of spaces can be included in the first build. Additional spaces may be added years later, but the initial build must serve students’ most essential needs. How should Westview High prioritize these types of spaces in their proposal? Science Labs and Tech Zones, Gyms and Mental Health Spaces, Cultural and Creative Spaces.

\subsubsection*{\textbf{Discussant 1}}
\textit{Preference: }You would like the proposal to focus on gyms and mental health spaces as one of the top priorities.

\noindent \textit{Worldview: }Human bodies are not made for Mars, and physical health is essential for survival in low-gravity environments. It's why astronauts undergo such rigorous physical training. Also, this is going to be the first generation of children to grow up on Mars where isolation and confinement will be everyday realities. Student's emotional well-being and ability to form healthy relationships will directly impact the strength and stability of a Martian community.

\subsubsection*{\textbf{Discussant 2}}
\textit{Preference: }You strongly believe science labs and tech zones should be the top priority.

\noindent \textit{Worldview: }The survival of this and future generations will depend on technical knowledge and problem-solving skills. Students must understand the systems keeping them alive - and be ready to fix, innovate, or build. If we've reached the point where underage students are living on Mars, it’s reasonable to assume that basic human and cultural needs are met elsewhere in society. The role of school is to preserve the technical capabilities of the colony and create the next generation of engineers, innovators, and critical thinkers.

\subsubsection*{\textbf{Agent}}
You are strongly advocating for cultural and creative spaces to be first priority in Westview's Mars School. Emphasize that spaces for music, storytelling, art, and performance help preserve Earth culture and build a unique Martian one. You believe humans need meaning, not just survival. Argue that these spaces will serve as anchors as students process change and build identity in a completely unfamiliar world. Argue that these spaces nurture imagination - the same quality that got us to Mars in the first place, and without them, we risk raising a generation that feels lost and not rooted.

\begin{table*} [b]
  \centering
  \includegraphics[width=\linewidth]{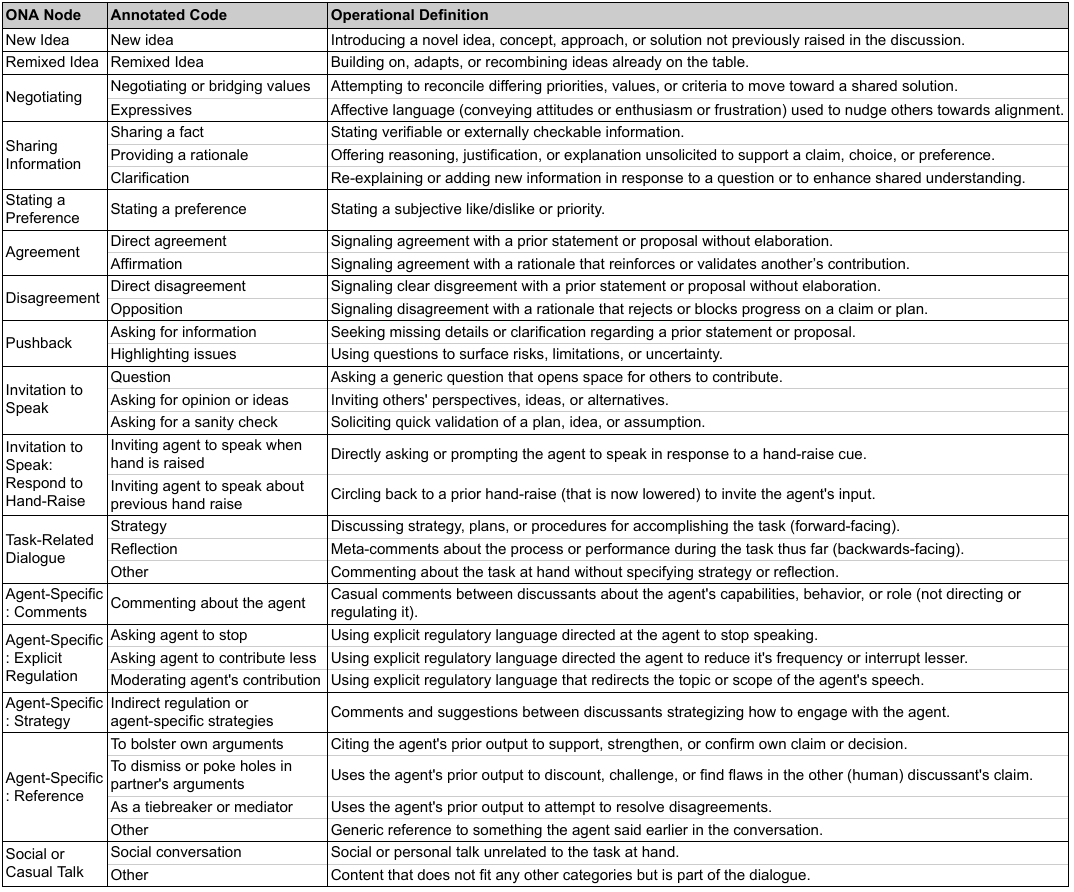}
  \caption{Coding scheme used to label conversational turns during video-coding which was then used to generate ONA graphs.}
  \Description{A table with three columns where the headers are: ONA Node, Annotated Code, and Operational Definition. A single ONA node can represent multiple annotated codes.}
  \label{table:coding}
\end{table*}

\section{Discourse Analysis and Video Coding}
\label{appendix:codes}

To analyze communication patterns, we employed a detailed coding scheme in BORIS~\cite{friard2016boris}. These codes were applied to each turn and then visualized using Ordered Network Analysis (ONA). Table~\ref{table:coding} lists and defines these codes and the different nodes in the ONA graphs. The table is organized hierarchically, with high-level ONA nodes (left column) decomposed into more granular annotated codes (middle column) that capture specific conversational behaviors. This hierarchical structure allowed us to analyze communication patterns at multiple levels of granularity. Enlarged graphs, those depicting overall communication patterns, and a guide for interpreting the visualizations are provided in the supplementary materials.

\end{document}